\UseRawInputEncoding
\documentclass[10pt,trans]{IEEEtran}
\usepackage{stfloats}
\usepackage{dingbat}
\usepackage{savesym}
\savesymbol{checkmark}
\usepackage{amsfonts}
\usepackage{amssymb}
\usepackage{amsthm}
\usepackage{cite}
\usepackage[colorlinks = true,
linkcolor = black,
urlcolor  = black,
citecolor = black,
anchorcolor = black]{hyperref}
\usepackage[cmex10]{amsmath}

\usepackage{float}
\usepackage{color}
\usepackage{algorithm}
\usepackage{algorithmic}
\usepackage{multirow}
\usepackage{graphicx}
\usepackage[normalem]{ulem}
\usepackage{tabularx}
\usepackage{subfigure}

\usepackage{fancybox,dashbox}
\usepackage{bbm}
\usepackage[font=small,justification=raggedright]{caption}
\usepackage[dvipsnames]{xcolor}
\usepackage{indentfirst}
\usepackage{graphicx}
\usepackage{tabularx,booktabs}
\usepackage{diagbox}
\usepackage{CJK}
\usepackage{threeparttable}
\usepackage{enumerate}
\newcolumntype{C}{>{\centering\arraybackslash}X}
\usepackage{svg}
\usepackage{pdfpages}
\usepackage{footmisc}
\usepackage{tablefootnote}

\begin{document}
	\title{Deep Learning for Wireless Networked Systems: a joint Estimation-Control-Scheduling Approach}
	\author{Zihuai Zhao, Wanchun Liu*,~\IEEEmembership{Member,~IEEE,} Daniel E.\ Quevedo,~\IEEEmembership{Fellow,~IEEE,}
		Yonghui Li,~\IEEEmembership{Fellow,~IEEE,} and Branka Vucetic~\IEEEmembership{Fellow,~IEEE}
		\thanks{Z.\  Zhao, W.\ Liu, Y.\ Li, and B.\ Vucetic are with School of Electrical and Information Engineering, The University of Sydney, Australia.
			Emails:	\{zihuai.zhao,\ wanchun.liu,\ yonghui.li,\ branka.vucetic\}@sydney.edu.au. 
			D.\ Quevedo is with the School of Electrical Engineering and Robotics, Queensland University of Technology (QUT), Brisbane, Australia.	Email: dquevedo@ieee.org. W. Liu is the corresponding author.}	
	}
	
\maketitle
\begin{abstract}
Wireless networked control system (WNCS) connecting sensors, controllers, and actuators via wireless communications is a key enabling technology for highly scalable and low-cost deployment of control systems in the Industry 4.0 era.
Despite the tight interaction of control and communications in WNCSs, most existing works adopt separative design approaches.
This is mainly because the co-design of control-communication policies requires large and hybrid state and action spaces, making the optimal problem mathematically intractable and difficult to be solved effectively by classic algorithms.
In this paper, we systematically investigate deep learning (DL)-based estimator-control-scheduler co-design for a model-unknown nonlinear WNCS over wireless fading channels.
In particular, we propose a co-design framework with the awareness of the sensor's age-of-information (AoI) states and dynamic channel states.
We propose a novel deep reinforcement learning (DRL)-based algorithm for controller and scheduler optimization utilizing both model-free and model-based data. An AoI-based importance sampling algorithm that takes into account the data accuracy is proposed for enhancing learning efficiency.
We also develop novel schemes for enhancing the stability of joint training. 
Extensive experiments demonstrate that the proposed joint training algorithm can effectively solve the estimation-control-scheduling co-design problem in various scenarios and provide significant performance gain compared to separative design and some benchmark policies.
\end{abstract}


\section{Introduction}\label{intro}
\begin{figure}[t]
	\centering
	\includegraphics[width=0.48\textwidth]{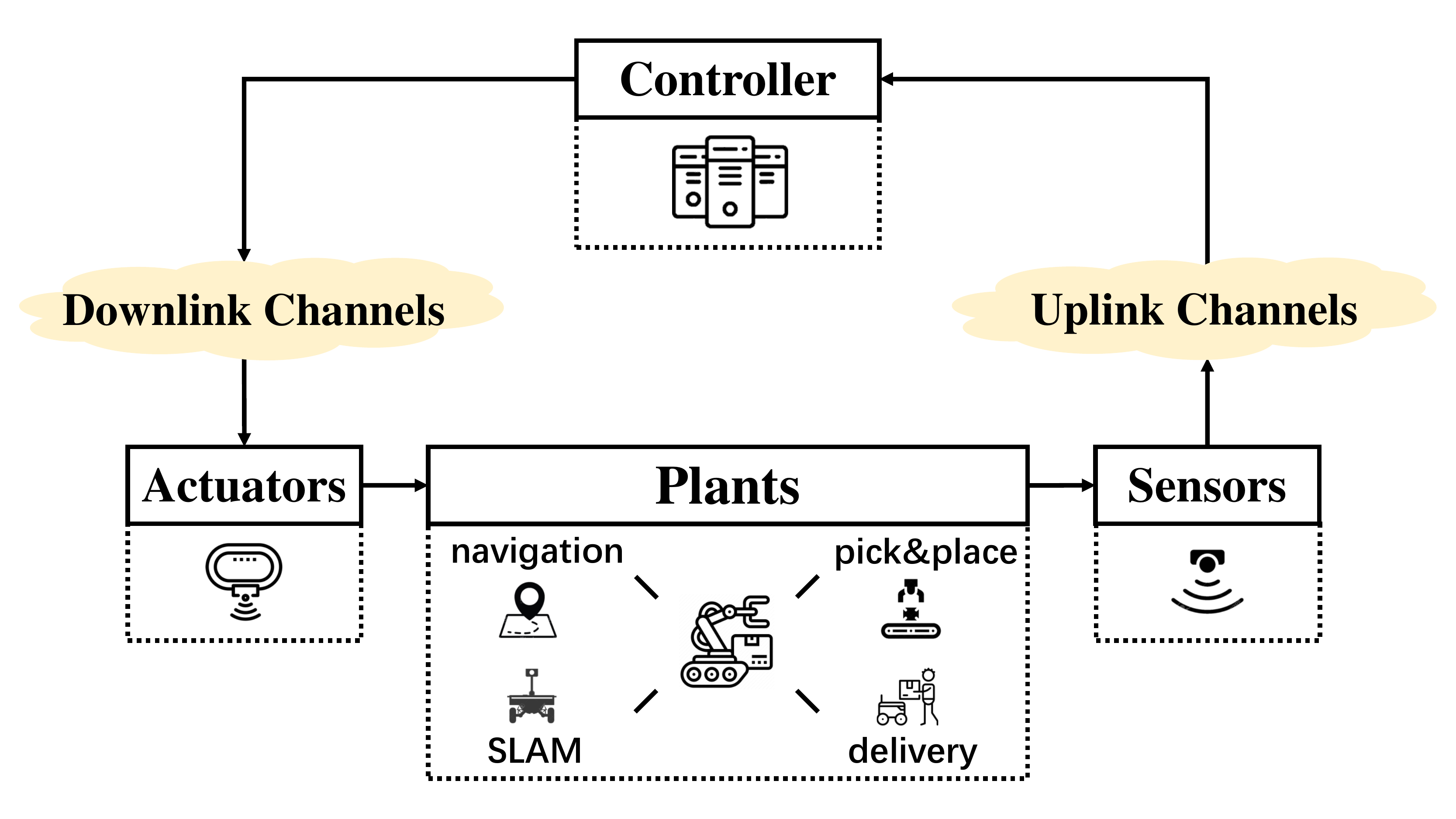}
	\caption{A wireless networked control system (WNCS).}    
	\label{General WNCS}  
\end{figure}

Under the rapid development of industrial applications in the Fourth Industrial Revolution, such as smart manufacturing, smart city, smart grids, e-commerce warehouses and industrial automation systems,  wireless networked control system (WNCS) has been considered as a key solution to the  high-scalable and low-cost deployment of ubiquitous automatic control systems~\cite{8166737}. A typical WNCS consisting of plants, sensors, actuators, and a controller is illustrated in Fig.~\ref{General WNCS}. In the feedback control loop of the WNCS, the sensors measure plant states and send them to the controller for processing and generating control signals via uplink channels, which will then be sent to the actuators for execution via downlink channels.

In principle, the nature of WNCS design is highly interdisciplinary, which involves signal processing for plant state estimation, control theory for optimally regulating the plant behavior, and communication theory for reliably transmitting the sensor and controller signals under limited communication resources.
Since both estimation and control rely on the information delivered by the communication system, the WNCS design for achieving the optimal control performance should jointly take into account the estimation, control, and communication algorithms that tightly interact with each other. Ideally, those algorithms should be jointly designed to optimize the control performance of WNCSs under resource constraints.

Although the concept of control-communication co-design in WNCSs was proposed decades ago (see \cite{8166737} and references therein), most related works from different research societies were built on the separative design principle.
The communications society focuses solely on improving the communications performance such as data rate, latency, and reliability, without taking into account the WNCS system dynamics, or performance~\cite{kountouris2021semantics, uysal2022semantic}. Although in the 5G era, ultra-reliable low latency communications have been proposed for mission-critical control applications, the prevailing design principle is standalone and not tailored to any control applications, where the control performance is not treated as a design objective~\cite{popovski2019wireless}.
On the other hand, the control system society's effort on WNCSs mainly focuses on the control (and estimation) algorithm design with predetermined communications policies (see \cite{schenato2007foundations} and its follow-up works).
The recent works on control-communication co-design for WNCSs can be categorized into \textbf{two streams}: \emph{control-aware communication design and control-communication policy co-design}.

In \textbf{Stream 1}, communication protocols are optimized to achieve the best control performance or under  certain control-related constraints. 
In \cite{gatsis2015opportunistic,Eisen}, transmission scheduling and power allocation problems of WNCSs were investigated for achieving the minimum overall transmission power consumption while guaranteeing certain control performance. 
In~\cite{EGWPeters}, a control-aware scheduler design problem was considered based on the communications protocol of IEEE 802.15.4.
In~\cite{huang2020wireless}, a communication protocol with variable packet length was proposed and optimized for achieving the best control performance.
In~\cite{liu2019wireless}, a transmission power allocation problem of a WNCS with a coding-free communication protocol was investigated, aiming to achieve optimal overall control performance.
In~\cite{liu2020latency}, a novel framework was developed for jointly optimizing the communication design parameters to achieve the best control performance.
Note that all those works are restricted to linear dynamical systems with linear control laws.
For remote state estimation of linear WNCSs, transmission scheduling problems have drawn significant attention. In \cite{Tomlin,HAN2017260,leong2017sensor,Wu2018Auto,LEONG2020108759,liu2021DRL}, optimal scheduling policies were investigated for various system setups to minimize average estimation errors.

In \textbf{Stream~2}, both the control and communication policies are jointly optimized to achieve the overall control performance. 
Stream~2 is more  challenging due to the fact that the joint policy has very large combined state and action spaces when taking into account both control and communications domains.
In a nutshell, most co-design problems can be formulated as dynamic decision-making ones. However, considering large state and action spaces, conventional solutions such as the Markov decision process cannot be applied due to the curse-of-dimensionality.
To solve this issue, most works in this stream rely on deep-learning (DL) approaches with artificial neural networks (NNs) for function approximations.
In~\cite{DRL2018CDC}, a deep reinforcement learning (DRL) approach was adopted to learn both the control and the transmission scheduling signals.
In particular, DRL combines artificial NNs with a framework of reinforcement learning that helps software agents learn how to solve decision-making problems and reach their goals. 
In~\cite{lima2020model}, both the control policy and the dynamic transmission power allocation policy were jointly optimized based on DRL.
It is worth noting that those DRL-based algorithms are model-free and are applied to the practical WNCS scenario that does not require accurate knowledge of the nonlinear system (plant) models, while the conventional solutions are purely model-based.

There are still many open problems in the area of \textbf{control-communications policy co-design with unknown nonlinear system models}. Many existing works, such as~\cite{DRL2018CDC,lima2020model}, assume that the sensor measurements are perfect and the (uplink) communications between sensor-controller are error-free. Under such an assumption, the controller has an accurate plant state in real time for generating control signals.
When considering a practical uplink channel, the controller does not always know the plant state and thus needs state estimation. This requires estimation-control co-design.
A key aspect is that the estimation quality significantly depends on the age of the sensor's information available to the estimator, which measures the time duration since the controller's last packet received from the sensor. 
Due to system dynamics and uncertainties, a larger age-of-information (AoI) of the sensor indicates a less reliable state estimate. 
For real-time control applications, an estimate with a small AoI is more important than the one with a large AoI. Such information about the data importance needs to be taken into account for the controller's training.
We note that the analysis and optimization of AoI in different communication networks have drawn significant attention during the past five years~\cite{yates2021age}. However, \textbf{how to leverage the AoI of sensor data for effectively training a controller has not been considered before}.
Furthermore, when considering DL-based estimator-control-communication co-design, one needs to systematically design a joint training algorithm for achieving time and performance efficiency, rather than training the three modules one by one. Otherwise, the resulting estimation, control and communication policies may not converge to desired ones, leading to poor overall control performance of the WNCS.
Due to aforementioned difficulties, \textbf{joint estimator-control-communication policy learning for WNCSs has not been investigated in the open literature}.

In this work, we systematically investigate a DL-based estimator-control-scheduler co-design framework for a model-unknown WNCS with nonlinear dynamic systems. We consider fading channels between sensor-controller and controller-actuator. The major contributions are summarized as follows.

\begin{itemize}
\item  We propose a novel DL-based WNCS over fading channels with time correlations. In particular, the AoI states of the sensor's information are utilized in the three modules of estimator, controller, and scheduler; both the controller and the scheduler leverage the fading channel states for decision-making. The instantaneous and historical states are utilized in each module.
Co-design frameworks for WNCSs with the awareness of AoI and channel states have not been considered in the open literature.

\item We develop a joint estimator-controller-scheduler training algorithm. In particular, we propose a DRL-based algorithm for controller and scheduler optimization utilizing both the model-free data that are received from the sensor directly and the model-based data that are generated by the estimator, when packet dropout occurs. An AoI-based importance sampling algorithm that takes into account the data accuracy is proposed for enhancing learning efficiency.
Moreover, we develop novel schemes for enhancing the stability of joint training.

\item Extensive experiments building on the OpenAI Gym platform demonstrate that the proposed joint training algorithm can effectively solve the estimation-control-scheduling co-design problem in various scenarios. Remarkable performance gains have been achieved compared to the separative design and some benchmark policies.
\end{itemize}

\textbf{Outline:} The system model of a general WNCS over fading channels is described in Section~\ref{problem formulation}. The estimation and control co-design problems of a low-mobility and a high-mobility WNCS were investigated in Sections~\ref{simple system} and~\ref{complex system}, respectively. The numerical results are demonstrated and discussed in Section~\ref{experiment}, followed by conclusions in Section~\ref{conclusion}.
	
	
\section{System Model}\label{problem formulation}
	
	\subsection{WNCS Model}\label{system model}
\begin{figure*}[t]
		\centering
		\includegraphics[width=0.65\textwidth]{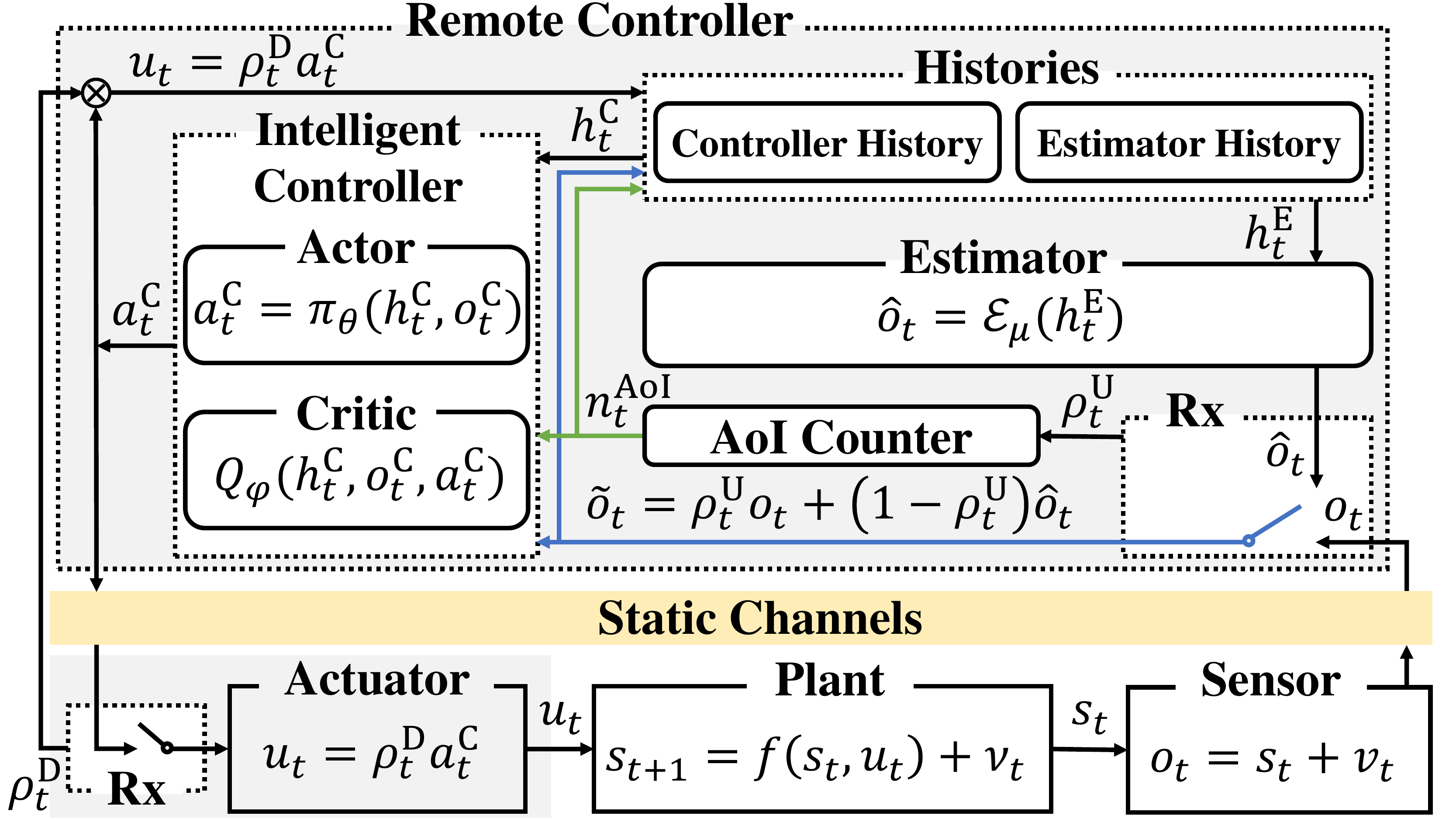}
		\caption{Estimator-controller co-design of the low-mobility WNCS.}    
		\label{Controller design}  
	\end{figure*}

	We consider a wireless networked control system  as shown in Fig.~\ref{Controller design}. The plant is a discrete-time  nonlinear system~as
	\begin{align}
		s_{t+1} &=  f(s_t,u_t) + \nu_t \label{plant dynamics}\\
		o_t &= s_t +  v_t\label{sensor dynamics}
	\end{align}
	where $s_t\in\mathbb{R}^{n_s}$ and $u_t\in\mathbb{R}^{n_u}$ are the plant state and the control input from the actuator at time $t$, respectively. In particular, the nonlinear dynamics $f:\mathbb{R}^{n_s}\times{\mathbb{R}^{n_u}}\to{\mathbb{R}^{n_s}}$ is unknown to the remote  controller, and $\nu_t$ is the plant disturbance.
	$o_t\in\mathbb{R}^{n_s}$ is the sensor measurement of the plant state $s_t$ affected by the measurement noise $v_t\in\mathbb{R}^{n_s}$.
	
	We model the uplink channel and the downlink channel as $m$-state Markov fading channels~
	\cite{sadeghi2008finite}. The channel states of the uplink and the downlink are denoted by $b_t^{\mathrm{U}}\in\mathcal{W}^{\mathrm{U}}\triangleq\{w_1^{\mathrm{U}},\dots,w_m^{\mathrm{U}}\}$ and $b_t^{\mathrm{D}}\in\mathcal{W}^{\mathrm{D}}\triangleq\{w_1^{\mathrm{D}},\dots,w_m^{\mathrm{D}}\}$, respectively. 
	Let $p_{i,j}^{\mathrm{U}}$ and $p_{i,j}^{\mathrm{D}}$ denote the channel state transition probabilities from state $i$ to $j$ of the uplink channel and the downlink channel, respectively, i.e.,
	\begin{equation}
		\begin{aligned}
			&p_{i,j}^{\mathrm{U}}\triangleq\textsf{Prob}[b_{t+1}^{\mathrm{U}}=w_j^{\mathrm{U}}|b_t^{\mathrm{U}}=w_i^{\mathrm{U}}],\\
			&p_{i,j}^{\mathrm{D}}\triangleq\textsf{Prob}[b_{t+1}^{\mathrm{D}}=w_j^{\mathrm{D}}|b_t^{\mathrm{D}}=w_i^{\mathrm{D}}].
		\end{aligned}
	\end{equation}
	Then, the channel state transition probability matrices of the channels are
	\begin{equation}\label{matrix 1}
		M^{\mathrm{U}}\triangleq
		\begin{bmatrix}
			p_{1,1}^{\mathrm{U}} & \dots  & p_{m,1}^{\mathrm{U}}\\
			\vdots & \ddots & \vdots\\
			p_{1,m}^{\mathrm{U}} & \dots  & p_{m,m}^{\mathrm{U}} 
		\end{bmatrix}
	\end{equation}
	and
	\begin{equation}\label{matrix 2}
		M^{\mathrm{D}}\triangleq
		\begin{bmatrix}
			p_{1,1}^{\mathrm{D}} & \dots  & p_{m,1}^{\mathrm{D}}\\
			\vdots & \ddots & \vdots\\
			p_{1,m}^{\mathrm{D}} & \dots  & p_{m,m}^{\mathrm{D}} 
		\end{bmatrix}.
	\end{equation}
	We assume that the instantaneous channel states, i.e., $b^{\mathrm{U}}_t$ and $b^{\mathrm{D}}_t$ are known to the controller by classical channel estimation schemes~\cite{goldsmith2005wireless}, while the dynamic channel models, i.e., $M^{\mathrm{U}}$ and $M^{\mathrm{D}}$ are not available.

	Let the binary variables $\rho^{\mathrm{U}}_{t}\in\{0,1\}$ and $\rho^{\mathrm{D}}_{t}\in\{0,1\}$ denote transmission failure and success of the uplink channel and the downlink channel at time $t$, respectively. The packet error probabilities at different channel states are
	\begin{equation}\label{SC dropout rate}
		d_i^{\mathrm{U}}\triangleq\textsf{Prob}[\rho^{\mathrm{U}}_{t}=0|b_t^{\mathrm{U}}=w_i^{\mathrm{U}}], \forall{i}\in\{1,\dots,m\} 
	\end{equation}
	and
	\begin{equation}\label{CA dropout rate}
		d_i^{\mathrm{D}}\triangleq\textsf{Prob}[\rho^{\mathrm{D}}_{t}=0|b_t^{\mathrm{D}}=w_i^{\mathrm{D}}], \forall{i}\in\{1,\dots,m\}.
	\end{equation} 
We assume that the actuator sends the one-bit acknowledge information $\rho^{\mathrm{D}}_t$ to the controller via a perfect feedback channel. This is a widely adopted assumption in wireless communications.

	\subsection{Control and transmission schedule} \label{est-ctrl-sch model}
	We consider both a \textbf{low-mobility} scenario (e.g., process control systems in factories) and a \textbf{high-mobility} scenario (e.g., unmanned aerial vehicles) of the WNCS. In the former scenario, the channel coherence time is much longer than each control time slot, and thus channel state is \emph{static}. We note that the system is stochastic in this scenario. Therefore, the Markov fading channels in~\eqref{matrix 1} and~\eqref{matrix 2} degrade to additive white Gaussian noise (AWGN) channels with constant channel states (i.e., $m=1$ in \eqref{SC dropout rate} and \eqref{CA dropout rate}). Due to the low mobility, sensors are often able to be connected to power grids, and the transmission power consumption is not a major concern. For the latter scenario, sensors are commonly powered by batteries. Due to the costly battery replacement operations, it is of significant interest to reduce the uplink transmission rate while guaranteeing a certain level of desired control quality.
	Therefore, an uplink transmission scheduler implemented at the controller will schedule the sensor's transmissions only when it is necessary. Let $a^{\mathrm{Tx}}_t \in \{0,1\}$ denote the scheduling action and  $\mathcal{S}(\cdot)$ denote the scheduling function mapping from input states to $a^{\mathrm{Tx}}_t$. We will discuss the input states in the following section.
	
	The control function $\mathcal{C}(\cdot)$ maps input states into the control signal $a^{\mathrm{C}}_t \in \mathbb{R}^{n_u}$.
	Considering the packet dropouts of the downlink channel, the control input is (see Fig.~\ref{Controller design})
	\begin{equation}
		u_t =\rho^{\mathrm{D}}_{t} a_t^{\mathrm{C}},
	\end{equation}
	and thus~\eqref{plant dynamics} can be rewritten as
	\begin{equation}\label{dynamics1}
		s_{t+1} = f(s_t,\rho^{\mathrm{D}}_{t}a_t^{\mathrm{C}})+ \nu_t.
	\end{equation}

	\subsection{Design Objectives and Challenges}\label{objectives}
	In the \emph{low-mobility scenario}, the control reward function of the WNCS at time $t$ depends on both the plant state and the control input as
	\begin{equation}\label{reward function}
		r_t \triangleq \mathcal{R}(s_t,u_t).
	\end{equation}
	Usually, the reward is large if $s_t$ is close to the desired plant state and the control input $u_t$ is small.
	Note that in the classical linear quadratic control scenario, the reward is $s_t^\top \Xi_s  s_t + u_t^\top \Xi_u  u_t$, where $\Xi_s$ and $\Xi_u$ are constant positive semi-definite matrices~\cite{schenato2007foundations}.
	Then, the long-term average performance of the WNCS is defined as
	\begin{equation}\label{eq:problem1}
		J \triangleq  \lim\limits_{T\rightarrow \infty} \sum_{t=1}^{T} \gamma^{(t-1)} r_t,
	\end{equation}
	where $\gamma\in (0,1)$ is a discount factor. A smaller $\gamma$ indicates that the future reward is less important.
	Thus, the optimal control problem is $\max_{\mathcal{C}(\cdot)} J$.
	
	In the \emph{high-mobility scenario}, the uplink transmission energy consumption and the scheduling policy should be taken into account. Let $\bar{e}$ denote the sensor transmission energy consumption each time.
	The reward function of the WNCS~is
	\begin{equation}
		r'_t \triangleq \mathcal{R}(s_t,u_t) - a^{\mathrm{Tx}}_t \bar{e}.
	\end{equation}
	Thus, the control-schedule co-design problem is
	\begin{equation}\label{eq:problem2}
		\max_{\mathcal{C}(\cdot), \mathcal{S}(\cdot)} J' \triangleq  \lim\limits_{T\rightarrow \infty} \sum_{t=1}^T \gamma^{(t-1)} r'_t.
	\end{equation}

	For the dynamic decision making problems~\eqref{eq:problem1} and \eqref{eq:problem2}, there are \textbf{several challenges}.
	First, the conventional decision-making problems require explicit model knowledge to apply existing dynamic programming algorithms, however, both the plant and channel dynamics are unknown.
	Second, existing decision-making problems, including both model-free and model-based ones, commonly assume that instantaneous rewards are available. However, due to the transmission scheduling and the packet dropouts of the uplink channel, the plant state may not be received by the remote controller each time. Thus, the instantaneous reward depending on the plant state is not always available. Note that classic partially observable Markov decision process (POMDP) problems assume that the instantaneous system states may not be obtained at all times, but the rewards are. Thus, the POMDP solutions cannot solve our problems.
	Last, the co-design problem~\eqref{eq:problem2} involves both the control policy $\mathcal{C}(\cdot)$ with continuous actions and the transmission schedule policy $\mathcal{S}(\cdot)$ with discrete actions, while most dynamic programming algorithms can only handle either discrete or continuous actions, not both.

	We will develop a novel DL-based framework to tackle the challenges above.
	The low-mobility and the high-mobility scenarios will be investigated in the sequel.

	\section{Estimation and Control Co-Design over Static Channels}\label{simple system}
	In this section, we consider a low-mobility WNCS with static uplink and downlink channels. 
	Considering the feature of the WNCS with uplink packet dropouts, we use a hybrid \textbf{model-based and model-free (MB-MF) data generation} algorithm for controller design:
	if a sensor packet is received, the measurement state is used directly by the controller for generating a control signal -- \textbf{the model-free part}; for time slots without sensor packets, the plant state is predicted by a DL-based model approximator, i.e.,  the estimator -- \textbf{the model-based part}.
	The controller is trained by a DRL algorithm, which takes into account the different accuracy levels of data generated by the MB-MF method.
	To enhance time efficiency, we propose to train both the DL-based estimator and the DRL-based controller together, rather than one agent after the other. Although the state estimation is inaccurate at the beginning, the DRL controller can take advantage of it to explore a wider range of actions.
	The system architecture of the estimator and the controller is shown in Fig.~\ref{Controller design}.

	We also note that a simpler \textbf{model-free} DRL method adopted in~\cite{8128911, meng2021memory}, which has no model-knowledge nor model approximator. Therefore, zero plant state is assumed when the sensor packet is unavailable at the controller, and the zero state is used for calculating the instantaneous control reward and is also sent to the controller for generating the control signal. Due to the poor estimation of the plant state and hence the reward, the method in~\cite{8128911, meng2021memory} cannot guarantee acceptable training performance in our scenario. We will provide numerical results to compare it with our proposed method in Section~\ref{experiment}.
	
	The detailed design of the estimator and the controller are given below.

	\subsection{DL-based Remote State Estimator}\label{Estimator section}
	If the sensor measurement $o_t$ is received by the remote controller, we can use $o_t$ to approximate the current plant state $s_t$ in \eqref{sensor dynamics}; otherwise, a model-based approach is needed, and we use a state estimator $\mathcal{E}(\cdot)$ with historical data for state prediction. Therefore, the estimated state can be written as 
	\begin{equation}\label{observation}
		\tilde{o}_t =  
		\begin{cases}
			o_t, &\text{ if } a^{\mathrm{Tx}}_t \rho^\mathrm{U}_t = 1 \\ 
			\hat{o}_t\triangleq \mathcal{E}(h_t^{\mathrm{E}}), &\text{ otherwise} 
		\end{cases}
	\end{equation}
	where $h_t^{\mathrm{E}}$ is an $\ell$-length historical data before time slot $t$ consisted of the estimated plant states and the control actions~as
	\begin{equation}
		h_t^{\mathrm{E}}\triangleq\begin{cases}
			(\tilde{o}_{t-\ell},u_{t-\ell},\dots,\tilde{o}_{t-1},u_{t-1}) & \parbox[t]{.1\textwidth}{if $t>\ell$}\\
			(0,0,\dots,\tilde{o}_{1},u_{1},\dots,\tilde{o}_{t-1},u_{t-1}) & \parbox[t]{.1\textwidth}{if $\ell\geq{t}>1$}\\
			(0,0,\dots,0,0) & \parbox[t]{.1\textwidth}{otherwise.}
		\end{cases}
	\end{equation}
	In particular, the estimator uses the previous $\ell$ estimates for current state prediction once a packet dropout occurs. We note that the history length $\ell$ is a hyper-parameter, and a larger $\ell$ can provide better performance in principle at the cost of higher estimator's computation complexity.
	
\begin{figure}[t]  
		\centering
		\includegraphics[width=0.31\textwidth, page=1]{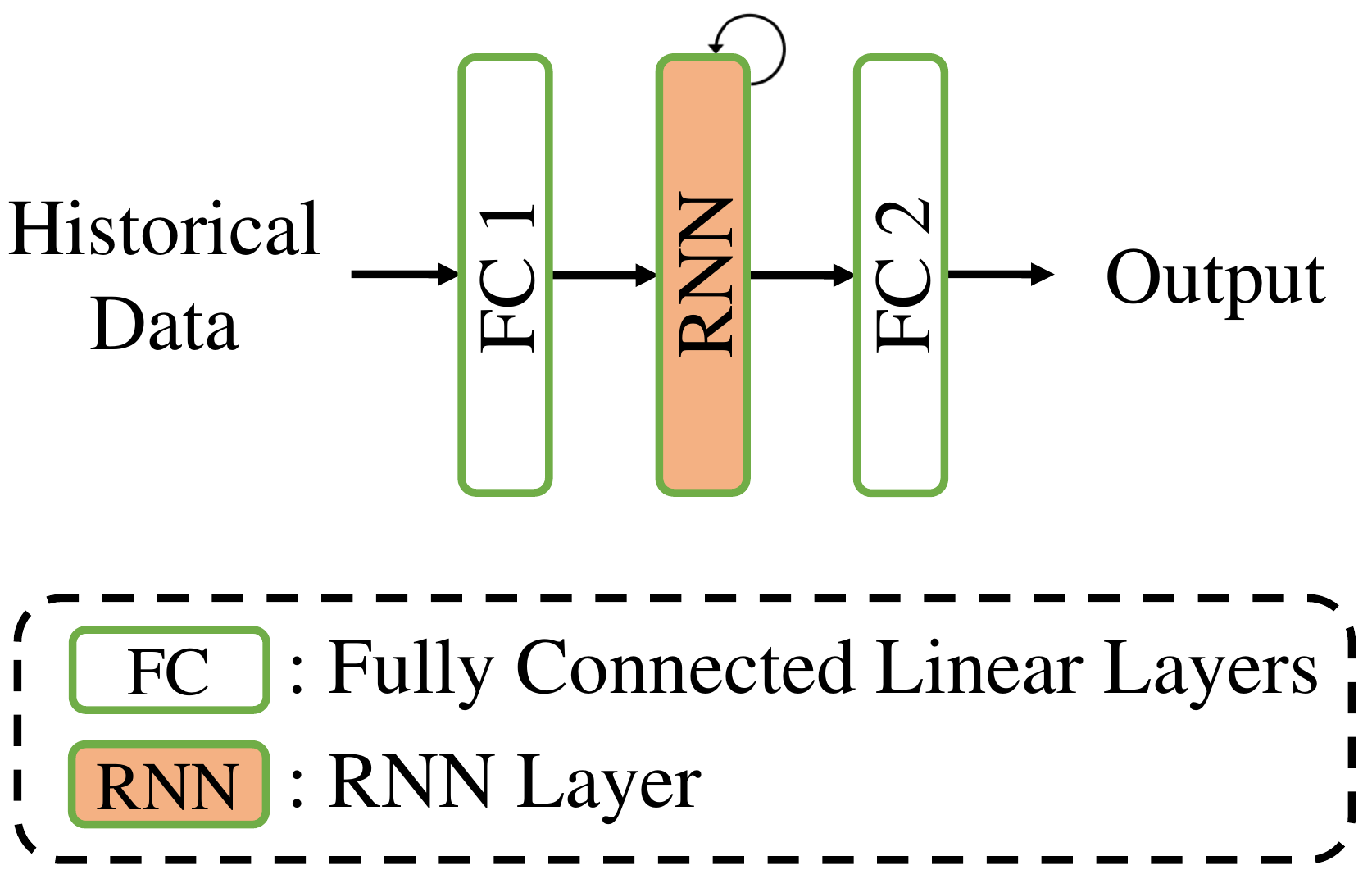}
		\caption{Network structure of the estimator.}    
		\label{estimator NN}  
	\end{figure}

	The estimator is approximated by a deep neural network (DNN) with parameter $\mu$, and is then denoted as $\mathcal{E}_\mu(\cdot)$.
	Compared with a vanilla feed-forward neural network (FFNN) with memory-less fully connected (FC) layers,
	a recurrent neural network (RNN) has connections between neurons that form a graph along a temporal sequence, allowing it to exhibit temporal dynamic behavior. Therefore, RNNs are more suitable for extracting long-term dependency features of multi-dimensional sequential inputs~\cite{7463717, sung2017learning, salmela2021predicting}.
	We use an RNN structure illustrated in Fig.~\ref{estimator NN}.
	
	The design problem of the estimator network is to minimize the difference between the real sensor measurements and the predicted states. Thus, the received observations $o_{t}$ and the corresponding histories $h^{\mathrm{E}}_{t}$ are used to train the estimator in a supervised learning fashion. Recall that the ground truth state $s_t$ is unavailable, and $o_t$ is an approximate of $s_t$.	
	For a batch of $N$ estimator training data $\{\mathcal{T}_i^{\mathrm{E}}\}_{i=1}^N\triangleq\{(h_{(i)}^{\mathrm{E}}, o_{(i)})\}_{i=1}^N$, where $i$ denotes the $i$th data in the batch, the loss function related to the expected estimation error is given by
	\begin{equation}\label{eq:est_loss}
		J_\mu = \frac{1}{N}\sum_{i=1}^N\|o_{(i)} - \mathcal{E}_\mu(h_{(i)}^{\mathrm{E}})\|^2.
	\end{equation}
	We use a gradient descent method to update the estimator NN for minimizing the loss function~\eqref{eq:est_loss}.
	
\subsection{DRL-based Controller with Hybrid Model-based and Model-free (MB-MF) Data}\label{hybrid control scheme}
The controller generates the control signal $a^{\mathrm{C}}_t$ based on the instantaneous MB-MF data $\tilde{o}_t$ and the history $h^{\mathrm{E}}_t$, i.e.,
\begin{equation}
	a^{\mathrm{C}}_t = \mathcal{C}(o_{t}^{\mathrm{C}} ,h^{\mathrm{C}}_t).
\end{equation}
where
\begin{equation}\label{controller input 1}
o_{t}^{\mathrm{C}} \triangleq (\tilde{o}_t, n_t^{\mathrm{AoI}}),
\end{equation}
and $n_t^{\mathrm{AoI}}\geq{0}$ denotes the AoI of $\tilde{o}_t$, which measures the time
elapsed since the latest sensor packet was received by the controller.
$h_t^{\mathrm{C}}$ is an $\ell$-length historical data including the past plant state estimates, the AoI states, and  the control inputs, which is given by
\begin{equation}
h_t^{\mathrm{C}}\triangleq\begin{cases}
(o_{t-\ell}^{\mathrm{C}},u_{t-\ell},\dots,o_{t-1}^{\mathrm{C}},u_{t-1}) & \parbox[t]{.1\textwidth}{if $t>\ell$}\\
(0,0,\dots,o_{1}^{\mathrm{C}},u_{1},\dots,o_{t-1}^{\mathrm{C}},u_{t-1}) & \parbox[t]{.1\textwidth}{if $\ell\geq{t}>1$}\\
(0,0,\dots,0,0) & \parbox[t]{.1\textwidth}{otherwise.}
\end{cases}
\label{controller history}
\end{equation}

Note that the AoI state indicates the accuracy of the current state estimate $\tilde{o}_t$. Such information is critical to the controller design. For example, when the AoI is large, the estimation is very inaccurate, and the controller may generate a zero control signal due to the largely unknown plant state.

We consider the Twin Delayed Deep Deterministic Policy Gradient (TD3)~\cite{fujimoto2018addressing}, a policy-based DRL algorithm for generating actions with a continuous space, to approximate and optimize the controller $\mathcal{C}(\cdot)$. 
In general, TD3 is a modification of the Deep Deterministic Policy Gradient (DDPG) algorithm~\cite{lillicrap2015continuous} to address the overestimation of the value estimate in actor-critic methods (e.g., DDPG). That is, since the actor is updated with respect to the maximization of the value estimate given by the approximated critic, the overestimated value estimate will cause a sub-optimal action to be highly rated by a sub-optimal critic. This leads to the sub-optimal action being reinforced in the next policy update, which creates a problematic feedback loop. In particular, TD3 introduces twin critic networks to reduce the overestimation.
It has shown significantly improved learning speed and performance compared to existing deterministic DRL algorithms with continuous action spaces over a variant of tasks~\cite{fujimoto2018addressing}. 
\emph{To obtain a tailored TD3 algorithm to our problem, we make two primary changes to the original one:}
First, we introduce RNNs into both the actor and the critic NNs for effectively processing the time-correlated historical data.
Second, we propose a novel AoI-based importance sampling method by properly taking into account the data accuracy to enhance the sampling efficiency.
	
\subsubsection{Network Structure of the DRL-based Controller}\label{OC section 1}
\begin{figure}[t]  
	\centering  
	\includegraphics[width=0.39\textwidth, page=1]{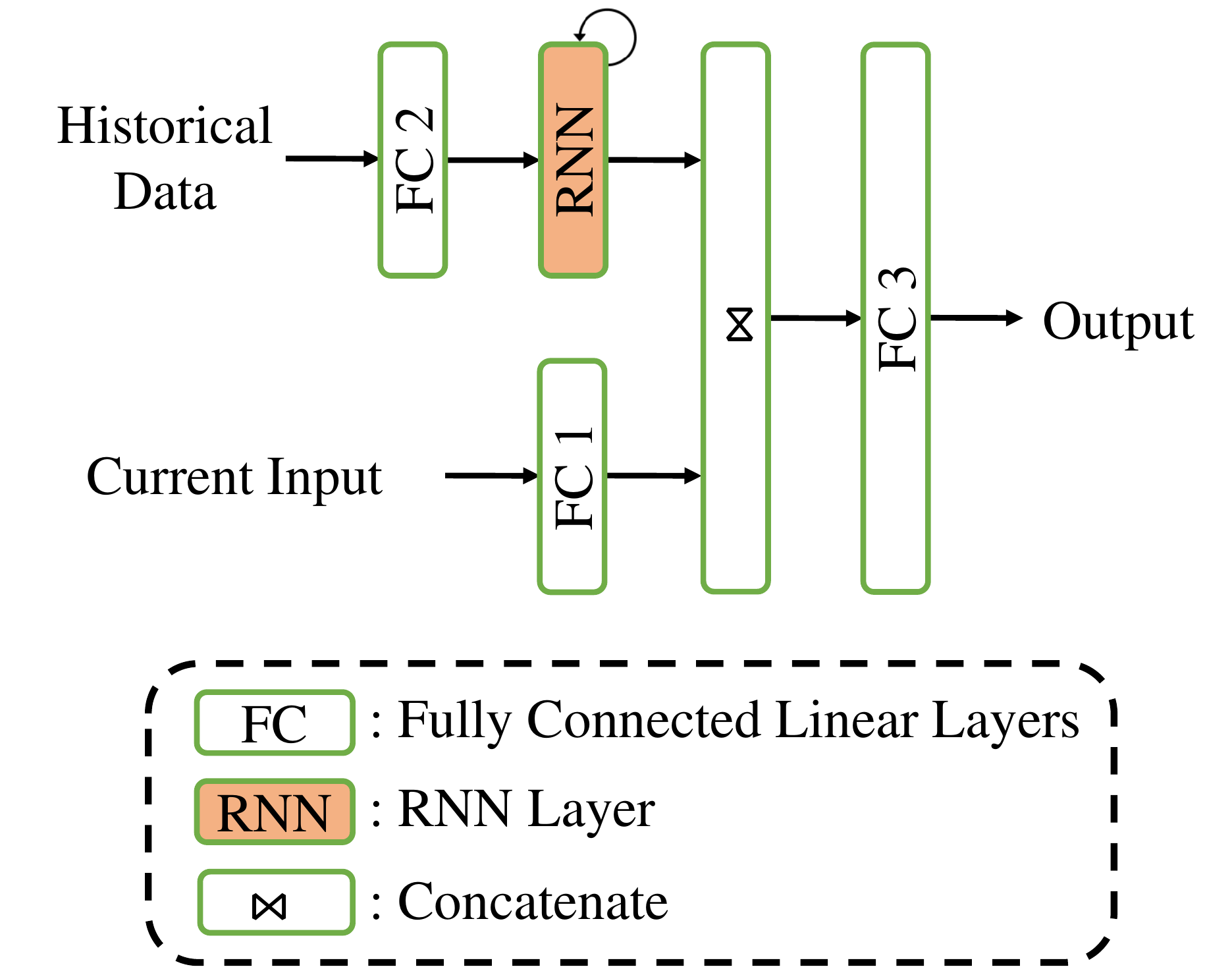}
	\caption{Network structure of the actor (and also the critic).}    
	\label{actor-critic NN}  
\end{figure}

The controller's actor and critic NNs have an identical network structure as illustrated in Fig.~\ref{actor-critic NN}. Note that the current input and the historical data are processed first separately, and an RNN is applied for processing the historical data with time correlations. 

\textbf{Actor network:} given the state input $(o_t^{\mathrm{C}},h_t^{\mathrm{C}})$, the output $a_t^{\mathrm{C}}$, and the network parameter set $\theta$, the actor NN approximates the control function as
\begin{equation}\label{actor}
a_t^{\mathrm{C}}=\pi_\theta(o_t^{\mathrm{C}},h_t^{\mathrm{C}}).
\end{equation} 

\textbf{Twin critic networks:} given the state input $(o_t^{\mathrm{C}},h_t^{\mathrm{C}})$ and the action input $a_t^{\mathrm{C}}$, the twin networks both approximate the Q-value, i.e., the expected long-term cost under policy $\pi_\theta$:
\begin{equation}
\begin{aligned}
&Q^{\theta}(o_t^{\mathrm{C}},h_t^{\mathrm{C}},a_t^{\mathrm{C}}) \\
&= \mathbb{E}\!\left[\!\sum_{t'=t}^{\infty}\!\gamma^{t'-t}r_{t'} \Big| o_t^{\mathrm{C}},h_t^{\mathrm{C}},a_t^{\mathrm{C}}, a_{t'}^{\mathrm{C}}=\pi_\theta(o_{t'}^{\mathrm{C}},h_{t'}^{\mathrm{C}}),\! \forall t'\!>\!\!t\!\right].
\end{aligned}
\end{equation}
The Q-value represents the long-term average performance of the current state-action pair, and a larger Q-value indicates a better control action. Thus, the Q-values will be used for optimizing the policy $\pi_\theta$.
Note that the twin critic NNs are only used for training. The trained actor NN is the only network used for deployment.

Let $\varphi_1$ and $\varphi_2$ denote the network parameters of the twin critic NNs.
The approximated Q-values are  $Q_{\varphi_1}( o_{t}^{\mathrm{C}}, h_{t}^{\mathrm{C}},a_{t}^{\mathrm{C}})$ and $ Q_{\varphi_2}(o_{t}^{\mathrm{C}}, h_{t}^{\mathrm{C}},a_{t}^{\mathrm{C}})$, respectively. To avoid overestimation of Q-values, where bad states are estimated as high values that can result in suboptimal policy updates and divergent behavior, only the smaller Q-value from the twin critic NNs is accepted for training the actor NN.

\subsubsection{Training of Intelligent Controller}\label{OC section 2}

We consider an off-policy training scheme for the actor-critic NNs, where transitions consisting of the current state, history, action, reward, and next state are collected and stored in a $N_{\mathrm{C}}$-length experience replay buffer, $\mathcal{D}_{\mathrm{C}}$.
During the policy update, transitions will be sampled from the replay buffer to simultaneously update the actor and critic NNs.

\textbf{Transitions in the replay buffer.} 
From the reward definition~\eqref{reward function}, the immediate reward depends on both the plant state $s_t$ and the control input $u_t$. 
We use the estimated plant state $\tilde{o}_t$ to approximate $s_t$.
Since the downlink (control) packet dropouts introduce randomness of the immediate reward, we use the expected reward for enhancing the smoothness and stability of the  controller training. Therefore, we have
\begin{equation}\label{avg reward}
\bar{r}_t\triangleq
\mathbb{E}_{\rho_t^{\mathrm{D}}}\left[\mathcal{R}(\tilde{o}_t,u_t)\right]
= (1-d^{\mathrm{D}})\mathcal{R}(\tilde{o}_t,a_t^{\mathrm{C}})+d^{\mathrm{D}}\mathcal{R}(\tilde{o}_t,0).
\end{equation}
Recall that $d^{\mathrm{D}}$ is the packet error probability of the downlink, and the control input is zero when a packet dropout occurs.
Thus, a transition is denoted as $<\!h_{t}^{\mathrm{C}}, o_{t}^{\mathrm{C}}, a^{\mathrm{C}}_{t}, \bar{r}_{t}, o_{t+1}^{\mathrm{C}}\!>$.
Then, the $i$th transition in the replay buffer is given as
\begin{equation}\label{controller transition}
	\mathcal{T}_i^{\mathrm{C}}\triangleq <\!h_{(i)}^{\mathrm{C}}, o_{(i)}^{\mathrm{C}}, a^{\mathrm{C}}_{(i)}, \bar{r}_{(i)}, o_{(i')}^{\mathrm{C}}\!>\in{\mathcal{D}_{\mathrm{C}}},
\end{equation}
where $o_{(i')}^{\mathrm{C}}$ denotes the next sampled state given $o_{(i)}^{\mathrm{C}}$ and $a^{\mathrm{C}}_{(i)}$.

\textbf{Actor and critic NN update with sampled transitions.} 
For each round of actor and critic NN update, the controller samples a batch of $N$ transitions from the reply buffer.
By taking into account the importance of different transitions, the optimal sampling probabilities are different. Let $P(i)$ denote the sampling probability of $\mathcal{T}^{\mathrm{C}}_i$, where $\sum_{i=1}^{N_{\mathrm{C}}} P(i) = 1$.
The importance sampling scheme will be discussed later in this section.

Let $\mathcal{T}_{\tilde{i}}^{\mathrm{C}}$ denote the $\tilde{i}$th sampled transition, which was the $\hat{i}$th transition in $\mathcal{D}_{\mathrm{C}}$ prior to the sampling. We then define the loss function for updating the actor NN as
\begin{equation}
J_\theta \triangleq \frac{1}{N}\sum_{{\tilde{i}}=1}^N w_{({\tilde{i}})}^{\mathrm{C}} Q_{\varphi_1}(h_{({\tilde{i}})}^{\mathrm{C}}, o_{({\tilde{i}})}^{\mathrm{C}}, a_{({\tilde{i}})}^{\mathrm{C}}),
\end{equation}
where 
$\omega_{({\tilde{i}})}^{\mathrm{C}}=\frac{1}{N_{\mathrm{C}}P(\hat{i})}$
is the importance-sampling weight of $\mathcal{T}_{\hat{i}}$.
Note that the weights aim to correct the bias introduced by the importance sampling, which changes the distribution of the sampled data~\cite{schaul2015prioritized}.
Then, we have the gradient for policy update as
\begin{equation}\label{actor update}
\nabla J_\theta = \frac{1}N\sum_{i=1}^{N} w_{(i)}^{\mathrm{C}} \nabla_\theta Q_{\varphi_1}(h_{(i)}^{\mathrm{C}}, o_{(i)}^{\mathrm{C}}, a_{(i)}^{\mathrm{C}})\nabla_\theta\pi_\theta(h_{(i)}^{\mathrm{C}}, o_{(i)}^{\mathrm{C}}).
\end{equation}

The loss function for updating the $k$th critic NN is based on the temporal-difference (TD) error, which is the difference of Q-value estimated by the current state-action pair and the next step pair.
In particular, the next step Q-value estimate is achieved by the Bellman equation as
\begin{equation}\label{target y1}
y_{(i)}^{\mathrm{C}} = \bar{r}_{(i)} + \gamma \min_{k=1,2}Q_{\varphi_k}(h_{(i')}^{\mathrm{C}}, o_{(i')}^{\mathrm{C}}, \pi_{\theta}(h_{(i')}^{\mathrm{C}}, o_{(i')}^{\mathrm{C}})).
\end{equation}
Recall that the smaller estimated Q-value in the next step is adopted to avoid overestimation.
Then, the TD-error of the $k$th critic NN is 
\begin{equation}
\mathsf{TD}_{k,i} \triangleq y_{(i)}^{\mathrm{C}} - Q_{\varphi_k}(h_{(i)}^{\mathrm{C}}, o_{(i)}^{\mathrm{C}}, a_{(i)}^{\mathrm{C}}), k \in \{1,2\}.
\end{equation}

Therefore, the loss function and the gradient of the $k$th critic NN are
\begin{equation}\label{eq:loss_TD}
J_{\varphi_k}= \frac{1}{N}\sum_{i=1}^{N} w_{(i)}^{\mathrm{C}} \left(\mathsf{TD}_{k,i}\right)^2,
\end{equation}
and
\begin{equation}\label{critic update}
\nabla J_{\varphi_k} = - \frac{2}{N} \sum_{i=1}^{N} \omega_{(i)}^{\mathrm{C}} \mathsf{TD}_{k,i} \nabla_{\varphi_k} Q_{\varphi_k}(h_{(i)}^{\mathrm{C}}, o_{(i)}^{\mathrm{C}}, a_{(i)}^{\mathrm{C}}),
\end{equation}
respectively.
Note that the loss functions $J_\theta$ and $J_{\phi_k}$ need to be maximized and minimized, respectively, by the gradient decent method.

\textbf{Novel importance sampling scheme.} 
Different from existing importance sampling schemes, the importance of each transition in our problem depends on \textbf{three aspects}: 1) the AoI indicating the accuracy of a transition, i.e., the one with small AoI should be sampled more often; 2) the freshness of the transition, i.e., an old transition should be sampled less and less often in the training; 3) the TD error, i.e., a transition with larger TD error should be sampled more often, as it is a more dominant term in \eqref{eq:loss_TD} determining how the critic NNs should be updated.
By properly taking into account the above factors, we propose a novel importance sampling scheme as below to enhance the sampling efficiency.
\begin{enumerate}
	\item \emph{Sampling probability}. The sampling probability of the $i$th transition $\mathcal{T}_i^{\mathrm{C}}\in{\mathcal{D}_{\mathrm{C}}}$ is given by
	\begin{equation}\label{IS prob}
	P(i)=\frac{\mathrm{rank}(i)^{-\alpha}}{\sum_{n=1}^{N_{\mathrm{C}}}\mathrm{rank}(n)^{-\alpha}}, 0<\alpha\leq1,
	\end{equation}
	where $N_{\mathrm{C}}$ is the size of the experience replay buffer $\mathcal{D}_{\mathrm{C}}$ and $\mathrm{rank}(i)$ denotes the rank of the transition $\mathcal{T}_i^{\mathrm{C}}$ in $\mathcal{D}_{\mathrm{C}}$, and the first transition in $\mathcal{D}_{\mathrm{C}}$ has the highest ranking. Thus, the highly ranked transitions (with small $\mathrm{rank}(i)$) are sampled with higher probabilities.
	\item \emph{New data insertion}. The most recently generated transition is inserted as the first one in $\mathcal{D}_{\mathrm{C}}$, and the rest of the buffer is shifted by one step. In this way, the new transitions are more likely to be highly ranked and sampled frequently.
	\item \emph{AoI and TD error-based periodic resorting}. The replay buffer is sorted every $n_{\mathrm{sort}}$ time step based on the ranking values of each transition.
	Let $V_i^{\mathrm{rank}}$ denote the ranking value of $\mathcal{T}_i^{\mathrm{C}}$. It should take into account both the accuracy of the data and the TD error. 
	In particular, since each transition contains two estimated plant states, the transition AoI is defined as
	\begin{equation}\label{transition AoI}
	I_i^{\mathrm{AoI}} = n_{(i)}^{\mathrm{AoI}} + n_{(i')}^{\mathrm{AoI}}.
	\end{equation}
	Since different transitions can have the same transition AoI, we should sort them based on the TD errors. Thus, the ranking value is defined as 	\begin{equation}\label{ranking value}
	V_i^{\mathrm{rank}}=-I_i^{\mathrm{AoI}}+2\left(\mathrm{sigmoid}\left((\mathsf{TD}_{1,i})^2 \right)-1/2\right),
	\end{equation}
	where the sigmoid-based function $2(\mathrm{sigmoid}(\cdot)-1/2)$ is to normalize the squared TD error in the range of $[0,1)$.
	A higher ranking value indicates an accurate transition with a large TD error, and the corresponding transition will be sorted with a smaller $\mathrm{rank}(i)$ and sampled more often.
	To reduce the computation complexity, we do not update all transitions' ranking values at each time step but only update the sampled ones.	
\end{enumerate}

The detailed DL-based algorithm for the estimation-control co-design of the low-mobility WNCS is presented in Algorithm~\ref{alg1}.
	
	\begin{algorithm}[!t]
		\caption{Deep learning-based estimation-control co-design for the low-mobility WNCS.}
		\label{alg1}
		\begin{algorithmic}[1]
			\renewcommand{\algorithmicrequire}{} 
			\STATE Initialize estimator network $\mathcal{E}_{\mu}$, actor network $\pi_{\theta}$, critic networks $Q_{\varphi_1},Q_{\varphi_2}$ with random parameters $\mu,\theta,\varphi_1,\varphi_2$
			\STATE Initialize target networks $\theta'\gets{\theta}, \varphi_1'\gets{\varphi_1}, \varphi_2'\gets{\varphi_2}$
			\STATE Initialize experience replay buffers $\mathcal{D}_{\mathrm{E}}$ and $\mathcal{D}_{\mathrm{C}}$ with sizes $N_{\mathrm{E}}$ and $N_{\mathrm{C}}$
			\FOR {$t=1,T$}
			\STATE \texttt{/* Receiving or Predicting */}
			\IF{$\rho^{\mathrm{U}}_{t}=1$}
			\STATE Receive observation $\tilde{o}_t=o_t$
			\ELSE
			\STATE Predict observation with estimator $\tilde{o}_t=\mathcal{E}_\mu(h_t^{\mathrm{E}})$
			\ENDIF
			\STATE Update AoI $n_t^{\mathrm{AoI}}$ to form actor input $o_t^{\mathrm{C}}\triangleq[\tilde{o}_t, n_t^{\mathrm{AoI}}]$
			\STATE \texttt{/* Interacting */}
			\STATE Select action with exploration noise $a_t^{\mathrm{C}}=\pi_\theta(h_t^{\mathrm{C}},o_t^{\mathrm{C}})+\epsilon,\epsilon\sim{\mathcal{N}(0,{\sigma_{\mathrm{expl}}}^2)}$ 
			\STATE Calculate expected reward $\bar{r}_t$ according to \eqref{avg reward}
			\STATE \texttt{/* Storing Transitions */}
			\STATE Store transition $\mathcal{T}_{i=t}^{\mathrm{C}}$ labeled with ranking value $V_i^{\mathrm{rank}} = -I_{i=t}^{\mathrm{AoI}}$ according to \eqref{transition AoI} as the first one in $\mathcal{D}_{\mathrm{C}}$
			\IF{$I_{i=t}^{\mathrm{AoI}}=0$}
			\STATE Store transition $\mathcal{T}_{j=t}^{\mathrm{E}}$ in $\mathcal{D}_{\mathrm{E}}$ 
			\ENDIF
            \STATE \texttt{/* Periodic Buffer Resorting */}
            \IF{$t$ mod $N_{\mathrm{C}}$}
            \STATE Sort $\mathcal{D}_{\mathrm{C}}$ according to \eqref{ranking value}
            \ENDIF
			\STATE \texttt{/* Updating Estimator */}
			\STATE Sample a mini-batch of $M$ transitions $\{\mathcal{T}_m^{\mathrm{E}}\}_{m=1}^M$ uniformly at random from $\mathcal{D}_{\mathrm{E}}$
			\STATE Update estimator $\mathcal{E}_\mu$ by $\mu \gets \mu - 
            {\alpha_\mu} {\nabla J_\mu}$ according to~\eqref{eq:est_loss} with learning rate $\alpha_\mu$
			\STATE \texttt{/* Updating Actor-critic */}
			\STATE Sample a mini-batch of $N$ transitions $\{\mathcal{T}_n^{\mathrm{C}}\}_{n=1}^N$ from $\mathcal{D}_{\mathrm{C}}$ according to \eqref{IS prob}
            \STATE Calculate the TD-errors $\{\mathsf{TD}_{1,n}\}_{n=1}^N$ of sampled transitions and update their ranking values in $\mathcal{D}_{\mathrm{C}}$ according to \eqref{ranking value}
			\STATE Update actor $\pi_\theta$ by $\theta\gets\theta + \alpha_\theta {\nabla J_\theta}$ according to~\eqref{actor update} with learning rate $\alpha_\theta$
			\STATE Update twin critics $Q_{\varphi_k} (k \in \{1,2\})$ by 
            ${\varphi_k} \gets {\varphi_k} - \alpha_\varphi{\nabla J_{\varphi_k}}$ according to~\eqref{critic update} with learning rate $\alpha_\varphi$
			\IF{$t$ mod $n_\mathrm{target}$}
			\STATE Update target networks:
			\STATE $\theta'\gets{\tau\theta + (1-\tau)\theta'}$
			\STATE $\varphi_1'\gets{\tau\varphi_1 + (1-\tau)\varphi_1'}$
			\STATE $\varphi_2'\gets{\tau\varphi_2 + (1-\tau)\varphi_2'}$
			\ENDIF
			\ENDFOR
		\end{algorithmic} 
	\end{algorithm}

\section{Estimation-Control-Scheduling Co-Design over Dynamic Fading Channels}\label{complex system}
\begin{figure}[t]
	\centering
	\includegraphics[width=0.48\textwidth]{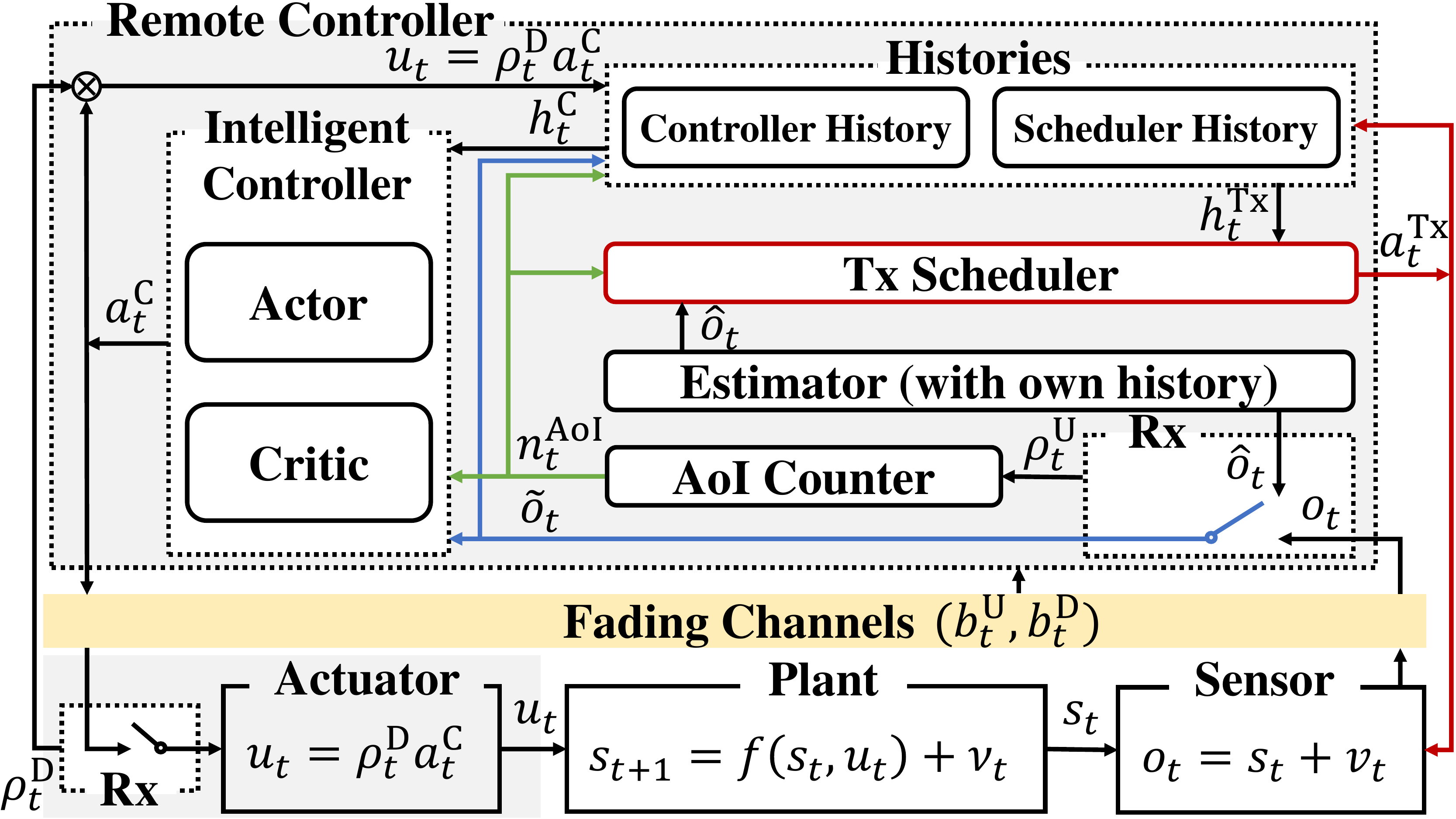}
	\caption{Estimator-controller-scheduler co-design of the high-mobility WNCS.}
	\label{Scheduler design}  
\end{figure}

In this section, we investigate the high-mobility WNCS over fading channels. Both the control performance and the sensor's transmission energy consumption are considered in the estimation-control-scheduler co-design, as illustrated in Fig.~\ref{Scheduler design}. 
We note that since the estimator is independent of the fading channel states and the scheduling policy, it is identical to the low-mobility scenario in Section~\ref{simple system}.
However, the controller is different due to the additional channel states.
Since the scheduler solves a dynamic decision-making problem, it is DRL-based, similar to the controller.
Due to the fact that the control and scheduling have continuous and discrete action spaces, respectively, they should be designed by different DRL algorithms.
Most importantly, the joint training of the two DRL agents brings new challenges. For example, the controller needs frequent sensor transmissions to have more accurate state estimation for high-quality control; while the scheduler might tend to transmit as less often as possible to reduce the transmission energy consumption.
It is critical to have a stable training process to achieve the desired tradeoff between them. 
In the following, we first present the details of the controller and the scheduler models, and then propose a novel DRL method for joint training.

\subsection{DRL-based Controller and Scheduler over Fading Channels}\label{scheduler section 1}
\begin{figure}[t]
	\centering
	\includegraphics[width=0.48\textwidth]{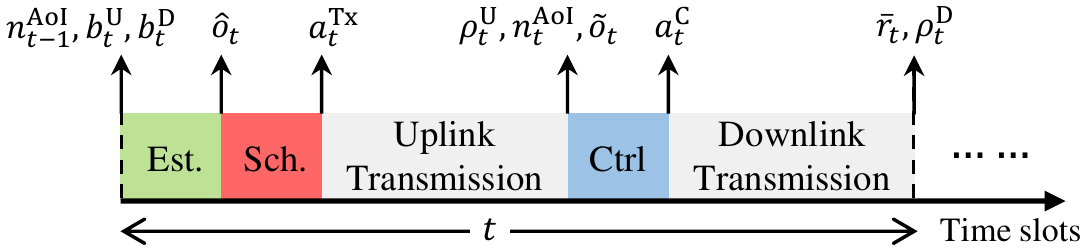}
	\caption{The sub-slots and the signals obtained after each sub-slot.}
	\label{fig:data_process}  
\end{figure}

Considering the state estimation, transmission scheduling, control signal generation, and uplink and downlink transmissions, each time slot can be divided into multiple sub-slots as shown in Fig.~\ref{fig:data_process} including the available signals after each sub-slot. Note that the scheduler and the controller have different plant state input (i.e., $\hat{o}_t$ and $\tilde{o}_t$, respectively) for decision making, since the scheduler operates before receiving the sensor's packet and can only utilize the state estimation $\hat{o}_t$ based on the previous sensor measurements.

The \textbf{controller} takes into account the estimated plant state, the AoI state, and the uplink and the downlink channel states for generating the control signal. The extended input of the controller is
\begin{equation}\label{controller input 2}
	o_{t}^{\mathrm{C}} \triangleq [\tilde{o}_t, b_t^{\mathrm{U}}, b_t^{\mathrm{D}}, n_t^{\mathrm{AoI}}].
\end{equation}
The action space and the actor-critic framework of the controller are identical to the low-mobility scenario.

The transmission \textbf{scheduler} makes decision based on the estimated plant state $\hat{o}_t$, the uplink channel state $b_t^{\mathrm{U}}$, and the AoI state  $n_{t-1}^{\mathrm{AoI}}$ from the previous time slot. Thus, the current state for the transmission schedule is defined as
\begin{equation}\label{scheduler input}
o_{t}^{\mathrm{Tx}}\triangleq[\hat{o}_t, b_t^{\mathrm{U}}, n_{t-1}^{\mathrm{AoI}}].
\end{equation}
Let  $h_t^{\mathrm{Tx}}$ denote an $\ell$-length state history of the scheduler as
\begin{equation}
h_t^{\mathrm{Tx}}\triangleq\begin{cases}
(o_{t-\ell}^{\mathrm{Tx}},a_{t-\ell}^{\mathrm{Tx}},\dots,o_{t-1}^{\mathrm{Tx}},a_{t-1}^{\mathrm{Tx}}) & \parbox[t]{.1\textwidth}{if $t>\ell$}\\
(0,0,\dots,o_{1}^{\mathrm{Tx}},a_{1}^{\mathrm{Tx}},\dots,o_{t-1}^{\mathrm{Tx}},a_{t-1}^{\mathrm{Tx}}) & \parbox[t]{.1\textwidth}{if $\ell\geq{t}>1$}\\
(0,0,\dots,0,0) & \parbox[t]{.1\textwidth}{otherwise}
\end{cases}
\label{scheduler history}
\end{equation}
The schedule decision at each time depends on both the state $o_{t}^{\mathrm{Tx}}$ and the history $h_t^{\mathrm{Tx}}$.

The scheduler adopts a Deep Q-Network (DQN)~\cite{mnih2015human} to generate actions since DQNs are commonly used for solving decision-making problems with discrete actions. Different from the actor-critic NNs, a DQN only approximates the Q-values at an input state with different actions.
Let $Q_{\phi}(\cdot)$ denote the DQN with parameter set $\phi$. Then, $Q_{\phi}(h_t^{\mathrm{Tx}}, o_t^{\mathrm{Tx}}, a_t^{\mathrm{Tx}})$ is the Q-value given the state input $(h_t^{\mathrm{Tx}}, o_t^{\mathrm{Tx}})$ and the action $a_t^{\mathrm{Tx}}$.
The scheduling action is given by
\begin{equation}\label{scheduler}
a_t^{\mathrm{Tx}}=\mathop{\arg\max}_{a \in \{0,1\}} Q_\phi(h_t^{\mathrm{Tx}}, o_t^{\mathrm{Tx}}, a),
\end{equation}
where the action leading to the highest Q-value is chosen.
Due to the historical input, we adopt the same RNN structure in Fig.~\ref{actor-critic NN} for the DQN.
We assume that the controller sends the one-bit scheduling signal $a_t^{\mathrm{Tx}}$ to the sensor via a perfect channel due to the negligible transmission overhead.

\subsection{Joint Training of Controller and Scheduler}\label{scheduler section 2}
By taking into account the sensor transmission energy consumption $\bar{e}$, the per-step overall reward for training the controller and scheduler is the control reward deducted by the energy consumption as
\begin{equation}\label{total performance}
	r_t^{\mathrm{total}}=\bar{r}_t - a^{\mathrm{Tx}}_t \bar{e},
\end{equation}
where the control reward $\bar{r}_t$ was defined in~\eqref{avg reward}.
As shown in Fig.~\ref{fig:data_process}, the overall reward can be calculated at the end of each time slot.

However, when training both the DRL-based controller and scheduler with the reward~\eqref{total performance}, \emph{convergence to desired policies is often difficult to achieve}.
This is mainly for two reasons:

1) At the beginning of the training, since the control reward $\bar{r}_t$ can be very low due to the lack of training of the controller, energy consumption is the dominant part of the overall reward. Thus, the scheduler tends not to schedule any sensor transmission. Note that any policy exploration, e.g., schedule of more sensor transmissions, will immediately lead to much lower overall rewards, especially when $\bar{e}$ is large.
Therefore, the controller can never be properly trained, and hence the control reward cannot be increased much, and the transmission energy consumption is always the dominant term. This inevitably leads to poorly trained control and scheduling policies.

2) When training the scheduler with the reward~\eqref{total performance}, since the scheduler's action has no direct (short-term) impact on the control reward term, it would always try to reduce the number of transmissions to achieve a higher overall reward. This wrong indention makes the joint controller and scheduler training inefficient, leading to poor performance. Furthermore, when the energy consumption $\bar{e}$ is large, the scheduler's stochastic action results in large variations of the overall reward, making the training process difficult to converge.

To solve the \textbf{first issue}, we propose to have a pre-training phase of the controller before the joint training, where the sensor is always scheduled for transmission. 
After the pre-training, the average control reward can be improved and comparable to the transmission energy consumption.

\emph{Training of the controller.}
The algorithm is identical to the low-mobility scenario in Section~\ref{hybrid control scheme}, where the only differences are to use the overall reward $r_t^{\mathrm{total}}$, instead of the control reward, and to operate with the scheduler's training at the same time. 

To solve the \textbf{second issue}, we should eliminate the impact of the instantaneous transmission energy cost on the scheduler's reward for the training purpose. Now, we propose to take the Q-value generated by the controller's DQN as the scheduler's reward, i.e.,
\begin{equation}\label{Q reward}
q_t = Q_{\varphi_1}(h_t^{\mathrm{C}},o_t^{\mathrm{C}},a_t^{\mathrm{C}}).
\end{equation}
This choice is motivated by the following two observations: 1) the controller's Q-value represents the expected long-term overall reward under the current control and scheduling policies, and a higher Q-value indicates a better schedule policy; and 2) the schedule action does not have a dominant instantaneous impact on the Q-value, which will avoid persistently reducing sensor transmissions.

\emph{Training of the transmission scheduler.}
Each transition for scheduler training is denoted as 
\begin{equation}\label{scheduler transition}
\mathcal{T}_i^{\mathrm{Tx}}\triangleq(h_{(i)}^{\mathrm{Tx}},o_{(i)}^{\mathrm{Tx}},a_{(i)}^{\mathrm{Tx}},q_{(i)},o_{(i')}^{\mathrm{Tx}})\in{\mathcal{D}_{\mathrm{Tx}}}
\end{equation}
where $\mathcal{D}_{\mathrm{Tx}}$ is the scheduler's reply buffer. Note that each scheduling action during training is selected based on the $\epsilon$-greedy policy.
We adopt the same importance sampling method as presented in Section~\ref{hybrid control scheme}. Given the importance-sampling weight $\omega_{(i)}^{\mathrm{Tx}}$, the loss function for optimizing the DQN is 
\begin{equation}\label{eq:sch_loss}
J_\phi \triangleq \frac{1}{N} \sum_{i=1}^{N} \omega_{(i)}^{\mathrm{Tx}}\left(\mathsf{TD}_i\right)^2
\end{equation}
where the TD error is defined as
\begin{eqnarray}
\mathsf{TD}_i \triangleq y_{(i)}^{\mathrm{Tx}} - Q_\phi(h_{(i)}^{\mathrm{Tx}}, o_{(i)}^{\mathrm{Tx}}, a_{(i)}^{\mathrm{Tx}}),
\end{eqnarray}
and 
\begin{equation}
y_{(i)}^{\mathrm{Tx}} = q_{(i)} + \gamma \max_{a}Q_{\phi}(h_{(i')}^{\mathrm{Tx}}, o_{(i')}^{\mathrm{Tx}}, a).
\end{equation}
Then, we have the gradient for DQN update as
\begin{equation}\label{actor update_DQN}
\nabla J_\phi =- \frac{2}N\sum_{\{\mathcal{T}_i^{\mathrm{C}}\}_{i=1}^N} \omega_{(i)}^{\mathrm{Tx}} \mathsf{TD}_i\nabla_\phi Q_{\phi}(h_{(i)}^{\mathrm{Tx}}, o_{(i)}^{\mathrm{Tx}}, a_{(i)}^{\mathrm{Tx}}).
\end{equation}

The details of the estimation-control-scheduler co-design in high-mobility WNCS are given in Algorithm~\ref{alg2}.
	
	\begin{algorithm}[!t]
		\caption{Deep learning-based estimation-control-scheduler co-design for the high-mobility WNCS.}
		\label{alg2}
		\begin{algorithmic}[1]
			\renewcommand{\algorithmicrequire}{} 
			\STATE Initialize estimator network $\mathcal{E}_{\mu}$, scheduler network $Q_{\phi}$, actor network $\pi_{\theta}$, critic networks $Q_{\varphi_1},Q_{\varphi_2}$ with random parameters $\mu,\phi,\theta,\varphi_1,\varphi_2$
			\STATE Initialize target networks $\phi'\gets{\phi}, \theta'\gets{\theta}, \varphi_1'\gets{\varphi_1}, \varphi_2'\gets{\varphi_2}$
			\STATE Initialize experience replay buffers $\mathcal{D}_{\mathrm{E}}$, $\mathcal{D}_{\mathrm{Tx}}$ and $\mathcal{D}_{\mathrm{C}}$ with sizes $N_{\mathrm{E}}, N_{\mathrm{Tx}}$ and $N_{\mathrm{C}}$
			\FOR {$t=1,T$}
			\STATE \texttt{/* Scheduling Transmission */}
			\STATE Predict observation with estimator $\hat{o}_t=\mathcal{E}_\mu(h_t^{\mathrm{E}})$ to form scheduler input $o_{t}^{\mathrm{Tx}}\triangleq[\hat{o}_t, b_t^{\mathrm{U}}, n_{t-1}^{\mathrm{AoI}}]$
			\STATE With probability $\epsilon_\mathrm{greedy}$ select a random action $a_t^{\mathrm{Tx}}\in{\{0,1\}}$ otherwise select action $a_t^{\mathrm{Tx}}$ according to~\eqref{scheduler}
			\STATE \texttt{/* Executing Control Signal */}
			\STATE Execute control signal $u_t =\rho^{\mathrm{D}}_{t} a_t^{\mathrm{C}}$ according to Algorithm~\ref{alg1} and generate Q-value $q_t = Q_{\varphi_1}(h_t^{\mathrm{C}},o_t^{\mathrm{C}},u_t)$ 
			\STATE \texttt{/* Storing Transitions */}
			\STATE Store transitions $\mathcal{T}_{i=t}^{\mathrm{C}}$ in $\mathcal{D}_{\mathrm{C}}$ and $\mathcal{T}_{j=t}^{\mathrm{E}}$ in $\mathcal{D}_{\mathrm{E}}$ according to Algorithm~\ref{alg1}
			\STATE Store transition $\mathcal{T}_{l=t}^{\mathrm{Tx}}$ labeled with ranking value $V_l^{\mathrm{rank}} = -I_{l=t}^{\mathrm{AoI}}$ according to \eqref{transition AoI} as the first one in $\mathcal{D}_{\mathrm{Tx}}$
            \STATE \texttt{/* Periodic Buffer Resorting */}
            \STATE Sort $\mathcal{D}_{\mathrm{C}}$ according to Algorithm~\ref{alg1}
            \IF{$t$ mod $N_{\mathrm{Tx}}$}
            \STATE Sort $\mathcal{D}_{\mathrm{Tx}}$ according to \eqref{ranking value}
            \ENDIF
			\STATE \texttt{/* Updating Estimator */}
			\STATE Update estimator $\mathcal{E}_\mu$ according to Algorithm~\ref{alg1}
			\STATE \texttt{/* Updating Actor-critic */}
			\STATE Update actor $\pi_\theta$, twin critics $Q_{\varphi_1}, Q_{\varphi_2}$ and target networks according to Algorithm~\ref{alg1}
			\STATE \texttt{/* Updating Scheduler */}
			\STATE Sample a mini-batch of $N$ transitions $\{\mathcal{T}_n^{\mathrm{Tx}}\}_{n=1}^N$ from $\mathcal{D}_{\mathrm{Tx}}$ according to \eqref{IS prob}
			\STATE Calculate the TD-errors $\{\mathsf{TD}_{1,n}\}_{n=1}^N$ of sampled transitions and update their ranking values in $\mathcal{D}_{\mathrm{Tx}}$ according to \eqref{ranking value}
			\STATE Update scheduler $Q_\phi$ by ${\phi} \gets {\phi} - \alpha_\phi{\nabla J_{\phi}}$ according to~\eqref{actor update_DQN} with learning rate $\alpha_\phi$
			\IF{$t$ mod $n_\mathrm{target}'$}
			\STATE Update target network $\phi'\gets{\phi}$
			\ENDIF
			\ENDFOR
		\end{algorithmic} 
	\end{algorithm}
	
\section{Numerical Experiments}\label{experiment}
In this section, we design simulations to evaluate the performance of the proposed co-design algorithms for the low-mobility and high-mobility WNCSs in Sections~\ref{simple system} and~\ref{complex system}, and compare them with some benchmark policies.
	
\subsection{Experiment Setups}
The sensor measurement noise is assumed to be Gaussian, i.e., $v_t\sim{\mathcal{N}(0,\sigma^2)}$. We set the history length as $\ell=3$. The communication parameters for the two scenarios are as below.
	
\subsubsection{Low-mobility WNCS}
We consider $6$ scenarios with different static packet error probabilities of the uplink channel and the downlink channel as well as different measurement noise powers, as shown in Table~\ref{tab:simple system}.
	
\subsubsection{High-mobility WNCS}
We model the uplink and downlink fading channels as two-state Markov chains, that $b_t^{\mathrm{U}}\triangleq\{w_1^{\mathrm{U}}, w_2^{\mathrm{U}}\}$ and $b_t^{\mathrm{D}}\triangleq\{w_1^{\mathrm{D}}, w_2^{\mathrm{D}}\}$, where both the states $w_1^{\mathrm{U}}$ and $w_1^{\mathrm{D}}$ have a packet error probability of $5\%$, and $w_2^{\mathrm{U}}$ and $w_2^{\mathrm{D}}$ have $10\%$.
We consider two channel state transition probability matrices $M_1$ and $M_2$ as
\begin{equation}
	M_1\triangleq
	\begin{bmatrix}
		0.7 & 0.3\\
		0.3 & 0.7
	\end{bmatrix}
\end{equation}
and
\begin{equation}
	M_2\triangleq
	\begin{bmatrix}
		0.3 & 0.7\\
		0.7 & 0.3
	\end{bmatrix},
\end{equation}
respectively. It is clear that the fading channel $M_1$ has a longer average channel state holding time, while $M_2$ leads to a more frequent change of channel states.
For the communication cost, we consider two scenarios with low communication cost $\bar{e}=5$ and high communication cost $\bar{e}=10$ respectively.
	
\begin{figure}[t] 
		\centering  
		\includegraphics[width=0.45\textwidth]{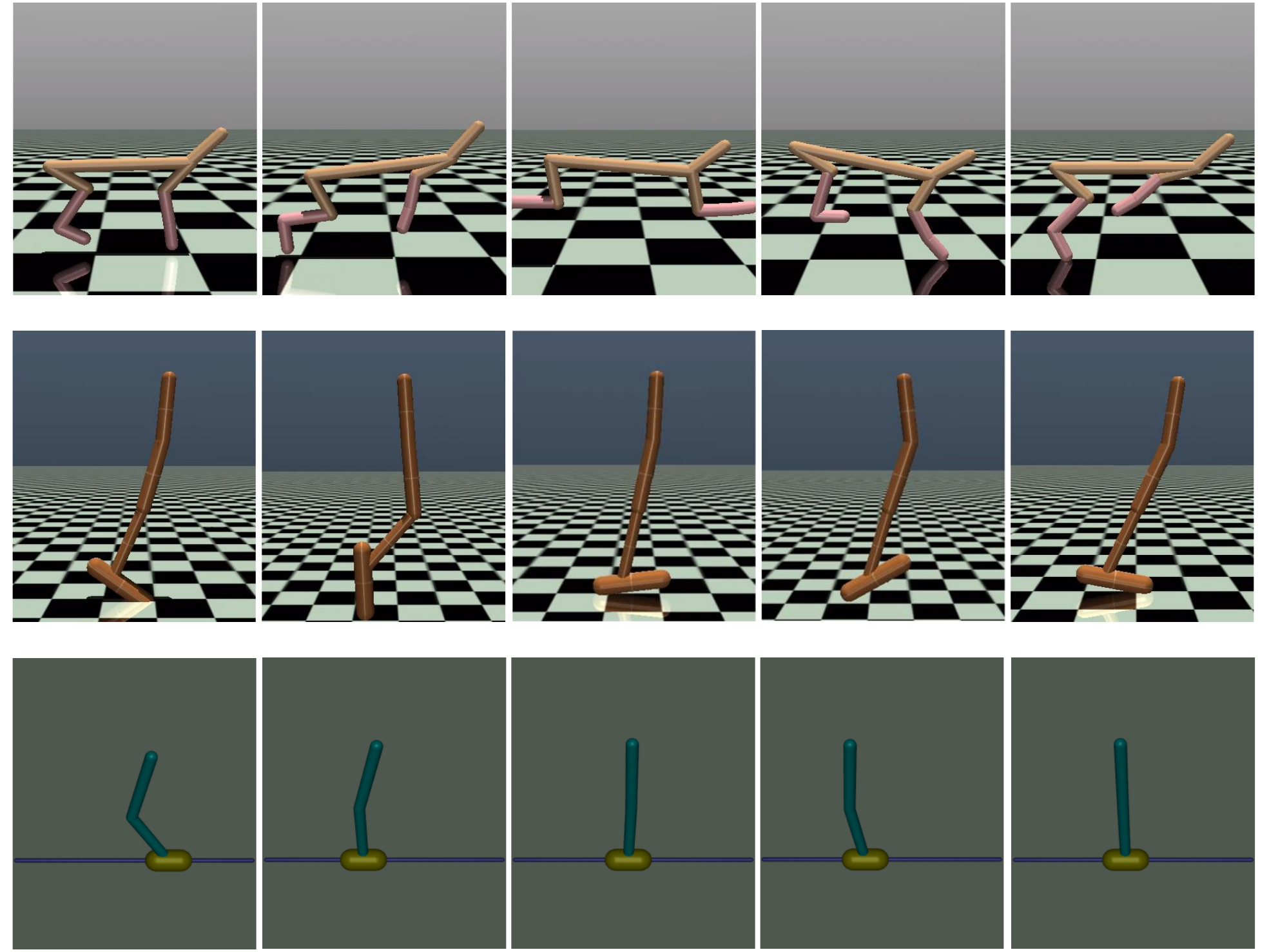}   
		\caption{Three MuJoCo tasks at different states: from top to bottom, are HalfCheetah, Hopper and InvDoublePen, respectively}    
		\label{MuJoCo}  
	\end{figure}

The plant system with unknown dynamics can be modeled by one of the MuJoCo tasks, well-known benchmarks for RL algorithms, in the OpenAI Gym\footnote{https://gym.openai.com/envs/\#mujoco} open-source simulation environment. We consider $3$ MuJoCo tasks: HalfCheetah-v2 (HalfCheetah), Hopper-v2 (Hopper) and InvertedDoublePendulum-v2 (InvDoublePen) to evaluate the performance of the proposed algorithms. A brief description and the state and control input dimensions of these MuJoCo tasks are given in Table~\ref{tab:Environment}, and the tasks with different states are illustrated in Fig.~\ref{MuJoCo}. The control performance of MuJoCo tasks is evaluated by the average sum of the control reward over a whole episode. A brief explanation of the control reward of the MuJoCo tasks is illustrated in Table~\ref{tab:control reward}. We note that the MuJoCo environment does not involve plant disturbance. Thus, we assume $\nu_t=0$ in our simulation.
	
	\begin{table}[t]
		\begin{threeparttable}
			\caption{Description of MuJoCo tasks}
			\label{tab:Environment}
			\setlength\tabcolsep{0pt} 
			
			\begin{tabular*}{\columnwidth}{@{\extracolsep{\fill}} lllc}
				\toprule
				\textbf{Plant} & \textbf{Description} & 
				\multicolumn{2}{c}{\textbf{Dimension}} \\ 
				\cmidrule{3-4}
				& & State & Ctrl Input \\
				\midrule
				HalfCheetah & Make a 2D cheetah robot run & 17 & 6 \\
				\midrule
				Hopper & Make a 2D 1-legged robot hop & 11 & 3 \\
				\midrule
				InvDoublePen & Balance 2-joint pole on a cart & 11 & 1 \\
				\bottomrule
			\end{tabular*}
			
			\smallskip
			\scriptsize
		\end{threeparttable}
	\end{table}
	
	\begin{table}[t]
		\begin{threeparttable}
			\caption{Description of control reward in MuJoCo}
			\label{tab:control reward}
			\setlength\tabcolsep{0pt} 
			
			\begin{tabular*}{\columnwidth}{l@{\hspace{2cm}}l}
				\toprule
				\textbf{Controlled plant} & \textbf{Reward formulation} \\ 
				\midrule
				HalfCheetah & FR - CC \\
				\midrule
				Hopper & FR + AR - CC \\
				\midrule
				InvDoublePen & AR - FR \\
				\bottomrule
			\end{tabular*}
			
			\begin{tablenotes}[flushleft]\footnotesize
				\item Forward reward (FR) depends on the displacement and velocity; Control cost (CC) depends on the magnitude of the control signal; Alive reward (AR) depends on the current state of the controlled plant \tablefootnote{https://mujoco.org/}.
			\end{tablenotes}
			
			\smallskip
			\scriptsize
		\end{threeparttable}
	\end{table}
	
	\begin{table*}[t]
		\caption{Comparison of control performance in the low-mobility scenario.}
		\label{tab:simple system}
		\begin{tabularx}{\textwidth}{@{}c*{7}{C}c@{}}
			\toprule
			\multicolumn{2}{c}{\textbf{Controlled plant}} & \multicolumn{3}{c}{\textbf{WNCS setup}} &  
			\multicolumn{3}{c}{\textbf{Control performance}} \\ 
			\cmidrule(lr){1-2}
			\cmidrule(lr){3-5}
			\cmidrule(lr){6-8}
			Name & Scenario label & Uplink channel dropout rate & Downlink channel dropout rate & Measurement noise $\sigma$ & Hybrid AoI & Hybrid Uniform & MF Uniform \\
			\midrule
			HalfCheetah  & 1 & 10\% & 10\% & 0.01 & ${6110 }\pm{368}$ & ${5306 }\pm{263}$ & ${4338 }\pm{145}$ \\ 
			& 2 & 10\% & 5\%  & 0.01 & ${6909 }\pm{167}$ & ${6263 }\pm{234}$ & ${5377 }\pm{253}$ \\ 
			& 3 & 5\%  & 10\% & 0.01 & ${7072 }\pm{203}$ & ${6627 }\pm{224}$ & ${6129 }\pm{206}$ \\ 
			& 4 & 10\% & 0\%  & 0.01 & ${8408 }\pm{397}$ & ${7511 }\pm{262}$ & ${6450 }\pm{373}$ \\ 
			& 5 & 0\%  & 10\% & 0.01 & ${8092 }\pm{306}$ & ${7555 }\pm{290}$ & ${7510 }\pm{292}$ \\
			& 6 & 10\% & 5\%  & 0.05 & ${6034 }\pm{325}$ & ${5393 }\pm{201}$ & ${4724 }\pm{331}$ \\ 
			\midrule
			Hopper       & 1 & 10\% & 10\% & 0.01 & ${2031 }\pm{177}$ & ${1774 }\pm{223}$ & ${1284}\pm{207}$ \\ 
			& 2 & 10\% & 5\%  & 0.01 & ${3102 }\pm{187}$ & ${2794 }\pm{246}$ & ${1825}\pm{354}$ \\ 
			& 3 & 5\%  & 10\% & 0.01 & ${2426 }\pm{267}$ & ${2118 }\pm{240}$ & ${1529 }\pm{359}$ \\ 
			& 4 & 10\% & 0\%  & 0.01 & ${3302 }\pm{179}$ & ${3015 }\pm{178}$ & ${2651 }\pm{297}$ \\ 
			& 5 & 0\%  & 10\% & 0.01 & ${2684 }\pm{261}$ & ${2517 }\pm{269}$ & ${2468 }\pm{273}$ \\
			& 6 & 10\% & 5\%  & 0.05 & ${2594 }\pm{301}$ & ${2144 }\pm{279}$ & ${1524 }\pm{199}$ \\ 
			\midrule
			InvDoublePen & 1 & 10\% & 10\% & 0.01 & ${1208 }\pm{146}$ & ${1144 }\pm{191}$ & ${949 }\pm{285}$ \\ 
			& 2 & 10\% & 5\%  & 0.01 & ${6647 }\pm{636}$ & ${5576 }\pm{766}$ & ${3832 }\pm{953}$ \\ 
			& 3 & 5\%  & 10\% & 0.01 & ${2203 }\pm{321}$ & ${1770 }\pm{215}$ & ${1324 }\pm{203}$ \\ 
			& 4 & 10\% & 0\%  & 0.01 & ${8844 }\pm{778}$ & ${8019 }\pm{839}$ & ${6294 }\pm{965}$ \\ 
			& 5 & 0\%  & 10\% & 0.01 & ${2555 }\pm{565}$ & ${2032 }\pm{417}$ & ${1990 }\pm{419}$ \\
			& 6 & 10\% & 5\%  & 0.05 & ${2506 }\pm{967}$ & ${1864 }\pm{439}$ & ${828 }\pm{133}$ \\ 
			\bottomrule
		\end{tabularx}
	\end{table*}
	
	\begin{table*}[t]
		\caption{Comparison of overall performance in the high-mobility scenario.}
		\label{tab:complex system}
		\begin{tabularx}{\textwidth}{@{}c*{7}{C}c@{}}
			\toprule
			\multicolumn{2}{c}{\textbf{Controlled plant}} & \multicolumn{3}{c}{\textbf{WNCS setup}} &  
			\multicolumn{3}{c}{\textbf{Overall performance}} \\ 
			\cmidrule(lr){1-2}
			\cmidrule(lr){3-5}
			\cmidrule(lr){6-8}
			Name & Scenario label & $M^{\mathrm{U}},M^{\mathrm{D}}$ & Communication cost & Measurement noise $\sigma$ & Scheduler-Q-Value & Scheduler-Reward & No-Scheduler \\
			\midrule
			HalfCheetah  & 7 & $M_1,M_1$ & 5 & 0.01 & ${2622}\pm{434}$ & ${1528}\pm{795}$ & ${2332}\pm{79}$ \\ 
			& 8 & $M_2,M_2$ & 5  & 0.01 & ${2374}\pm{393}$ & ${1192}\pm{762}$ & ${2133}\pm{90}$ \\ 
			& 9 & $M_1,M_1$  & 10 & 0.01 & ${-1316}\pm{624}$ & ${-1705}\pm{926}$ & ${-2793}\pm{84}$ \\ 
			& 10 & $M_2,M_2$ & 10  & 0.01 & ${-1539}\pm{710}$ & ${-2219}\pm{1367}$ & ${-3126}\pm{87}$ \\ 
			\bottomrule
		\end{tabularx}
	\end{table*}

The NN parameters of the estimator, controller, and scheduler are given in Table~\ref{tab:NN layers}. The hyper-parameters for NN training are provided in Table~\ref{tab:Hyper-parameters}. 
	
	\begin{table*}[t]
		\caption{NN parameters.}
		\label{tab:NN layers}
		\begin{tabularx}{\textwidth}{@{}l*{15}{C}c@{}}
			\toprule
			\textbf{NN Layers:} & \multicolumn{2}{|@{}c@{\hskip0.25in}|}{\textbf{FC 1}} & 
			\multicolumn{2}{@{}c@{\hskip0.25in}|}{\textbf{FC 2}}  &
			\multicolumn{4}{@{}c@{\hskip0.25in}|}{\textbf{FC 3}}  &
			\multicolumn{2}{@{}c@{\hskip0.25in}|}{\textbf{RNN}} \\ 
			\midrule
			\textbf{NN type:} & \multicolumn{2}{@{}c@{\hskip0.25in}}{\text{FFNN}} & 
			\multicolumn{2}{@{}c@{\hskip0.25in}}{\text{FFNN}}  &
			\multicolumn{2}{@{}c@{\hskip0.25in}}{\text{FFNN 1}}  &
			\multicolumn{2}{@{}c@{\hskip0.25in}}{\text{FFNN 2}}  &
			\multicolumn{2}{@{}c@{\hskip0.25in}}{\text{GRU}} \\ 
			\midrule
			\textbf{Parameters:} & DIM & AF & DIM & AF & DIM & AF & DIM & AF & DIM & AF  \\ 
			\midrule
			\textbf{Estimator} & $[n_{\mathrm{his}}^{\mathrm{E}},128]$ & ReLU & $[128,n_{\mathrm{out}}^{\mathrm{E}}]$ & Linear & $\backslash$ & $\backslash$ & $\backslash$ & $\backslash$ & $[128,128]$ & Linear\\ 
			\textbf{DQN} & $[n_{\mathrm{cur}}^{\mathrm{Tx}},128]$ & ReLU & $[n_{\mathrm{his}}^{\mathrm{Tx}},128]$ & ReLU & $[256,128]$ & ReLU & $[128,2]$ & Linear & $[128,128]$ & Linear\\ 
			\textbf{Actor} & $[n_{\mathrm{cur}}^{\mathrm{C}},128]$ & ReLU & $[n_{\mathrm{his}}^{\mathrm{C}},128]$ & ReLU & $[256,128]$ & ReLU & $[128,n_{\mathrm{out}}^{\mathrm{C}}]$ & Tanh & $[128,128]$ & Linear\\ 
			\textbf{Critic} & $[n_{\mathrm{cur}}^{\mathrm{C}}+n_{\mathrm{out}}^{\mathrm{C}},128]$ & ReLU & $[n_{\mathrm{his}}^{\mathrm{C}},128]$ & ReLU & $[256,128]$ & ReLU & $[128,1]$ & Linear & $[128,128]$ & Linear\\ 
			\bottomrule
		\end{tabularx}
		
		\hfill
		\\ The parameters row presents the input-output dimension (DIM) and the activation function (AF) of each NN layer. The $n_{\mathrm{cur}}^{\mathrm{C}}, n_{\mathrm{cur}}^{\mathrm{Tx}}$ denote the dimensions of $o_t^\mathrm{C}, o_t^\mathrm{Tx}$ and $n_{\mathrm{his}}^{\mathrm{E}}, n_{\mathrm{his}}^{\mathrm{C}}, n_{\mathrm{his}}^{\mathrm{Tx}}$ denote the dimensions of $h_t^\mathrm{E}, h_t^\mathrm{C}, h_t^\mathrm{Tx}$ respectively, and $n_{\mathrm{out}}^{\mathrm{E}}, n_{\mathrm{out}}^{\mathrm{C}}$ are the dimensions of $o_t, a_t^{\mathrm{C}}$ respectively. 
		The parameters $n_{\mathrm{cur}}^{\mathrm{C}}, n_{\mathrm{cur}}^{\mathrm{Tx}}, n_{\mathrm{his}}^{\mathrm{E}}, n_{\mathrm{his}}^{\mathrm{C}}, n_{\mathrm{his}}^{\mathrm{Tx}}, n_{\mathrm{out}}^{\mathrm{E}}, n_{\mathrm{out}}^{\mathrm{C}}$ in the MuJoCo tasks are given as follows: HalfCheetah: $20, 19, 69, 78, 60, 17, 6$; Hopper: $14, 13, 42, 51, 42, 11, 3$; InvDoublePen: $14, 13, 36, 45, 42, 11, 1$.
	\end{table*}
	
	\begin{table}[t]
		\begin{threeparttable}
			\caption{Hyper-parameters for NN training}
			\label{tab:Hyper-parameters}
			\setlength\tabcolsep{0pt} 
			
			\begin{tabular*}{\columnwidth}{@{\extracolsep{\fill}} ll}
				\toprule
				\textbf{Hyper-parameter} & \textbf{Value} \\ 
				\midrule
				Size of replay buffers $N_{\mathrm{E}},N_{\mathrm{C}},N_{\mathrm{Tx}}$ & $10^{5}$\\
				Batch size $M, N$ & $100$ \\
				History length $\ell$ & $3$ \\
				\midrule
				Discount factor $\gamma$ & $0.99$ \\
				Scheduler $\epsilon$-greedy probability $\epsilon_\mathrm{greedy}$ & 0.1 \\
				Actor exploration noise $\sigma_{\mathrm{expl}}$ & 0.1 \\
				Target update delay $n_\mathrm{target}$ & $2$ \\
				Target update delay $n_\mathrm{target}'$ & $100$ \\
			Target update rate $\tau$ & $0.005$ \\
				\midrule
				Optimizer type & Adam \\
				Estimator learning rate $\alpha_\mu$ & $10^{-3}$ \\
				Scheduler learning rate $\alpha_\phi$ & $3\times{10^{-4}}$ \\
				Actor learning rate $\alpha_\theta$ & $3\times{10^{-4}}$ \\
				Critic learning rate $\alpha_\varphi$ & $3\times{10^{-4}}$ \\
                \midrule
                Sampling probability prioritization $\alpha$ & $1$ \\
				\bottomrule
			\end{tabular*}
			
			\smallskip
			\scriptsize
		\end{threeparttable}
	\end{table}
	
\subsection{Performance Evaluation and Comparison}

\subsubsection{Joint vs. separative estimation-control methods of the low-mobility scenario.}	
We consider scenarios 2 and 3 of the InvDoublePen task in Table~\ref{tab:simple system} to compare the control performance between the joint and separative estimation-control methods.\footnote{The other scenarios show the same trend as the selected ones and thus are omitted for brevity.} For a separative estimation-control method, the estimator will first be trained by using an independent and identically distributed (i.i.d.) random control inputs, after which the trained estimator will be adopted to assist the intelligent controller training according to Section~\ref{simple system}. We evaluate three setups of the separately training method. 

\begin{figure}[t]
	\centering
	\includegraphics[width=0.48\textwidth, page=1]{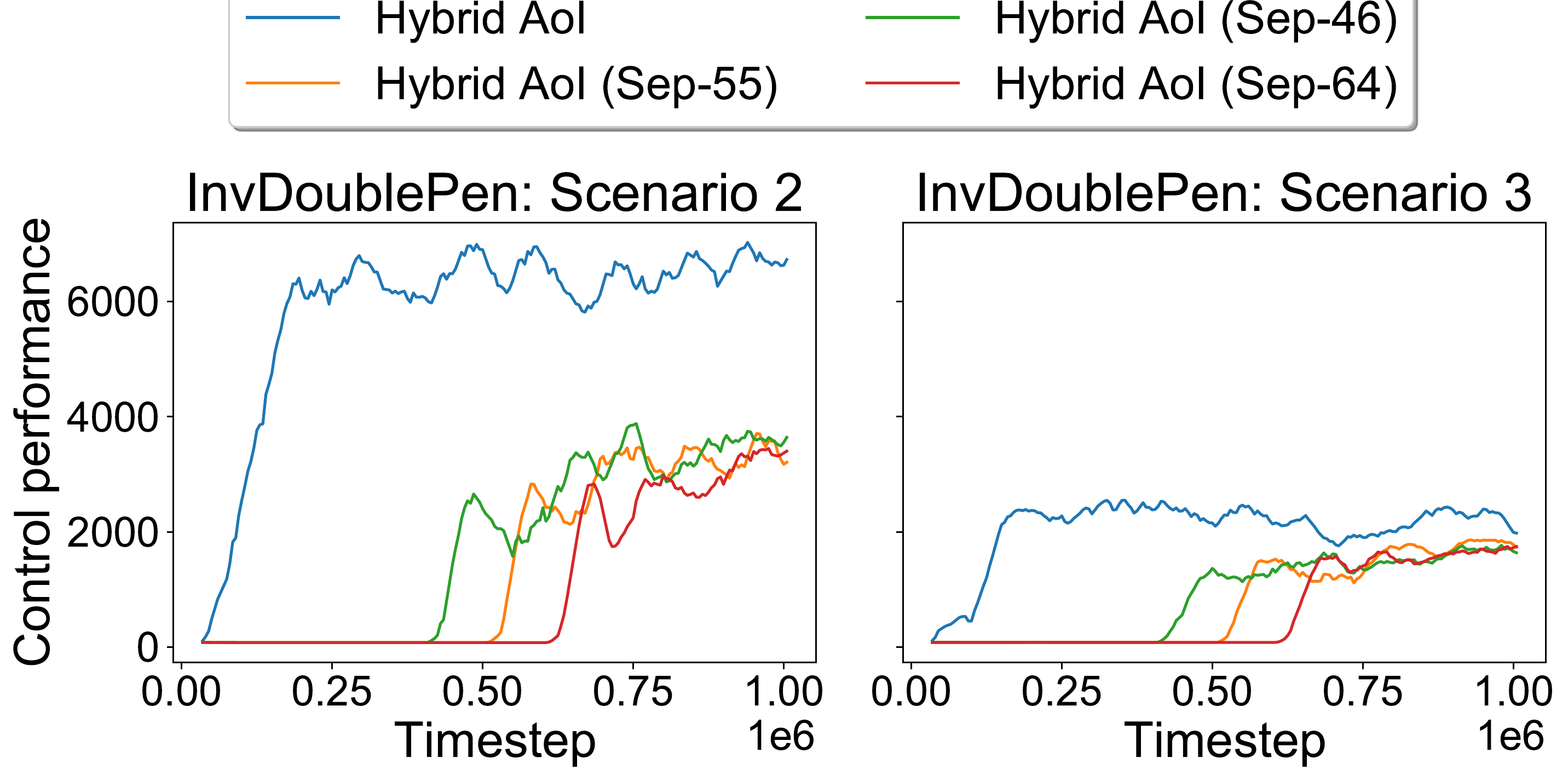}
	\caption{Comparison of the learning curves of the joint and separative estimation-control methods. Hybrid AoI (in blue) denotes the proposed joint algorithm and the remaining three are the separative ones  (see Table~\ref{tab:Compared methods 1} for explanation).}
	\label{fig:learning curve0}
\end{figure}

We plot the learning curves of the joint and separative training methods in Fig.~\ref{fig:learning curve0}. We see that the control performance of the joint method during the training improves stably and is significantly higher than the separative ones. For example, the control performance doubled when completing the training in Scenario 2. This is because during the separative training of the estimator, the i.i.d. random control inputs are not able to explore the state-action space of the plant effectively. Thus, the DL-based estimator can only learn part of the plant dynamics and provide inaccurate estimations when training the controller. The proposed joint method explores the plant dynamics more efficiently for estimator training, which in turn enhances the controller's optimization.

\subsubsection{Performance evaluation of the co-design algorithm in the low-mobility scenario}
in Section~\ref{simple system}, we propose a novel DRL framework for controller training with the hybrid MB-MF experience replay buffer and the AoI-based importance sampling method. To verify the effectiveness of the proposed algorithm, we compare it (Hybrid AoI) with two baseline algorithms, i.e., the one (Hybrid Uniform) with the hybrid MB-MF replay buffer and the uniform data sampling, and the one (MF Uniform) with a purely MF replay buffer (no plant state estimation) and the uniform sampling, as described in Table~\ref{tab:Compared methods 1}. 

\begin{table}[t]
	\begin{threeparttable}
		\caption{Benchmark Policies (low-mobility scenario).}
		\label{tab:Compared methods 1}
		\setlength\tabcolsep{0pt} 
		
		\newcommand{\tabincell}[2]{\begin{tabular}{@{}#1@{}}#2\end{tabular}}  
		
		\begin{tabular*}{\columnwidth}{@{\extracolsep{\fill}} ll}
			\toprule
			\textbf{Method} & \textbf{Description} \\ 
			\midrule
			Hybrid AoI (Sep-55) & \tabincell{l}{Hybrid MB-MF controller with AoI-based \\ importance sampling, where the estimator and the \\ controller are trained separately with \\ $5 \times 10^5$ time steps each.} \\
			\midrule
			Hybrid AoI (Sep-46) & \tabincell{l}{Hybrid MB-MF controller with AoI-based \\ importance sampling, where the estimator and the \\ controller are trained separately with \\ $4 \times 10^5$ and $6 \times 10^5$ time steps, respectively.} \\
			\midrule
			Hybrid AoI (Sep-64) & \tabincell{l}{Hybrid MB-MF controller with AoI-based \\ importance sampling, where the estimator and the \\ controller are trained separately with \\ $6 \times 10^5$ and $4 \times 10^5$ time steps, respectively.} \\
			\midrule
			Hybrid AoI (proposed) & \tabincell{l}{Hybrid MB-MF controller with AoI-based \\ importance sampling, where the estimator and the \\ controller are jointly trained.} \\
			\midrule
			Hybrid Uniform & \tabincell{l}{Hybrid MB-MF controller with uniform random \\ importance sampling, where the estimator and the \\ controller are jointly trained.} \\
			\midrule
			MF Uniform & \tabincell{l}{MF controller with uniform random experience  \\ replay (i.e., the controller receives zero-valued \\ observations when packet dropout occurs).} \\
			\bottomrule
		\end{tabular*}
		
		\smallskip
		\scriptsize
	\end{threeparttable}
\end{table}
	
\begin{figure*}[t]
		\centering
		\includegraphics[width=\textwidth, page=1]{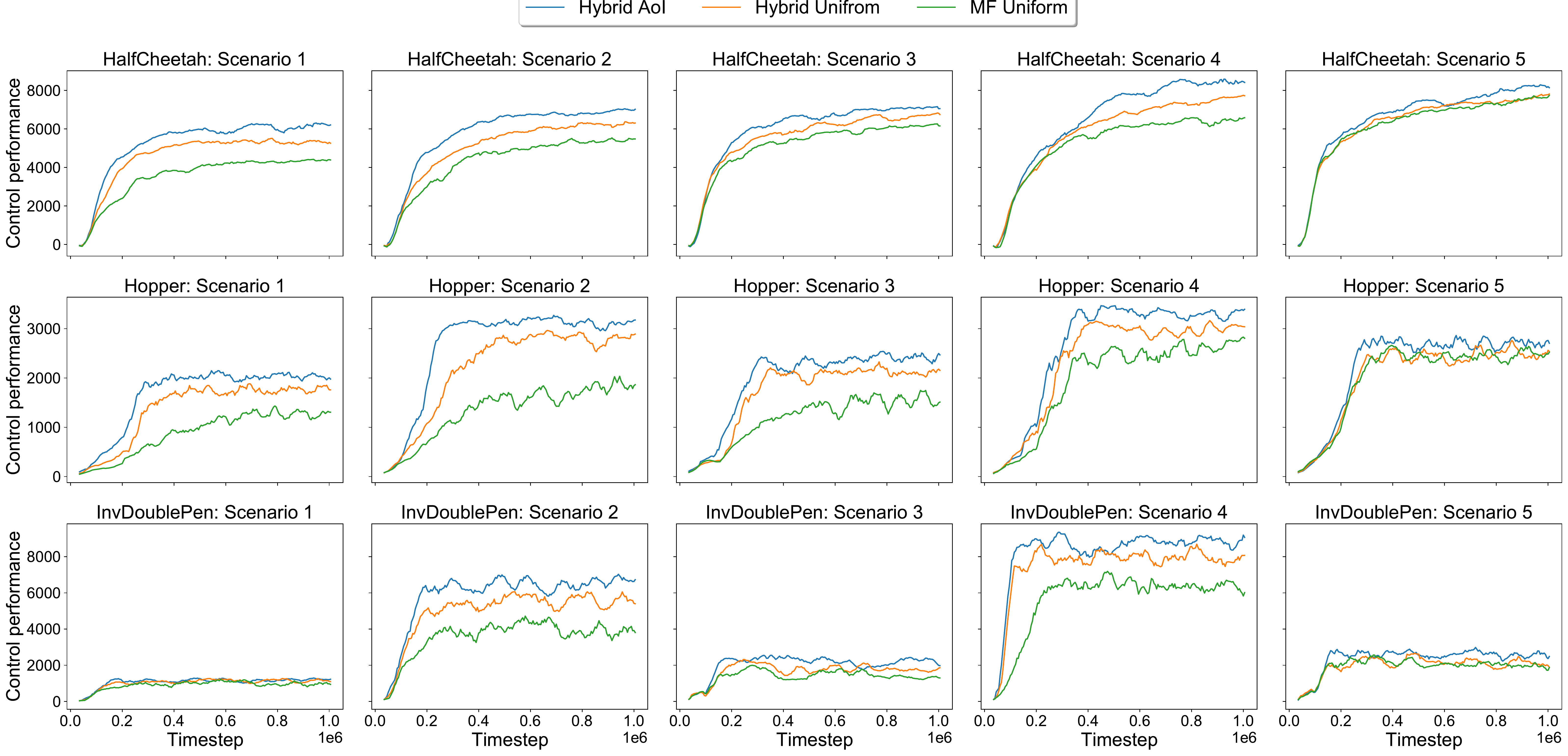}
		\caption{Comparison of learning curves in the low-mobility scenario - Part 1: different channel conditions.}
		\label{fig:learning curve1}
	\end{figure*}

\begin{figure}[t]
		\centering
		\includegraphics[width=0.49\textwidth, page=1]{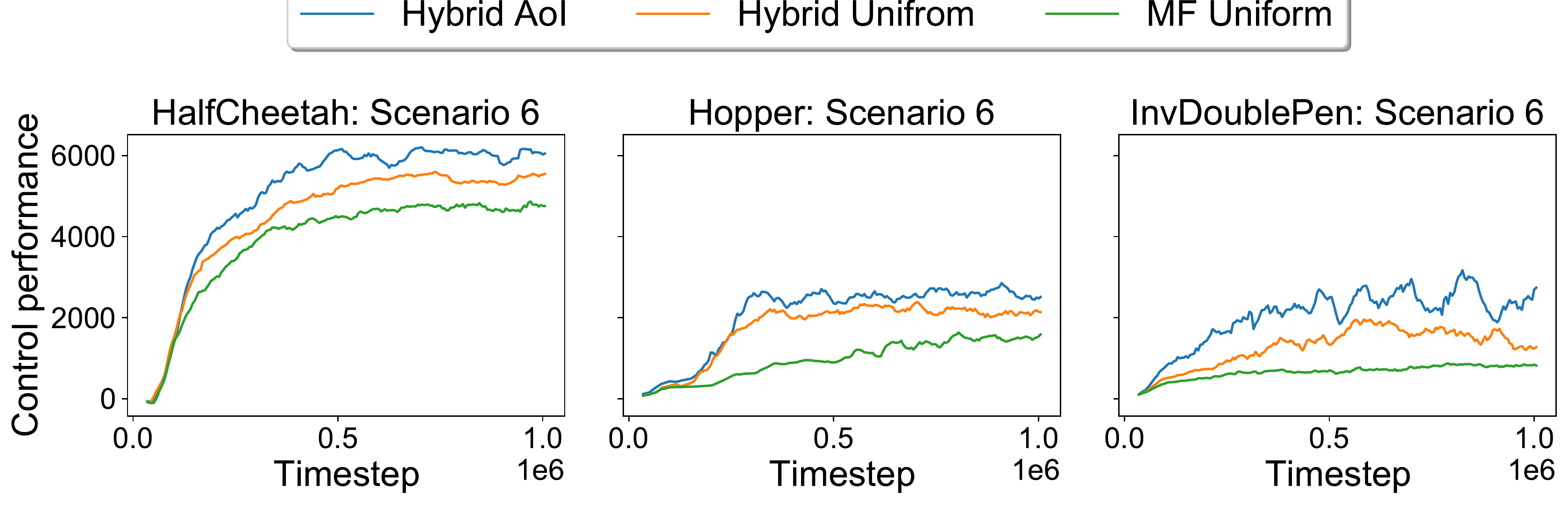}
		\caption{Comparison of learning curves in the low-mobility scenario - Part 2: different measurement noises.}
		\label{fig:learning curve2}
	\end{figure}

\textbf{Performance evaluation of the proposed algorithm over three MuJoCo tasks:} 
Figs.~\ref{fig:learning curve1} and~\ref{fig:learning curve2} show the learning performance of the proposed algorithm and the baseline ones of three control tasks over six different WNCS settings (see Table~\ref{tab:simple system} for explanations). 

We see that the InvDoublePen suffers most from high packet error probability and measurement noise while the HalfCheetah takes the minimum impact, e.g., when comparing scenario 1 (less reliable communication) with scenario 2 (more reliable communication). This is because the InvDoublePen is the most difficult control task due to the fact that the cart (i.e., the yellow part in the last row of Fig.~\ref{MuJoCo}) of the InvDoublePen is the only thing one can control for balancing the double-joint inverted pendulum. The task is impossible in the presence of relatively high packet error probability and measurement noise. On the other hand, for the HalfCheetah task, as shown in Table~\ref{tab:control reward}, the control reward does not consist of an alive reward, which means the HalfCheetah is more robust than the others and thus becomes less sensitive to packet dropouts and measurement noises.
	
\textbf{Comparison between hybrid MB-MF and MF methods.} 
From Figs.~\ref{fig:learning curve1} and~\ref{fig:learning curve2}, it is clear that the proposed algorithm has the best learning performance.
In particular, when comparing the two baseline algorithms, the hybrid MB-BF replay buffer-based approach outperforms the conventional MF one in all scenarios. In scenario 2 of the Hopper task, a $50\%$ control performance improvement has been achieved.

Besides the learning curves, we have included the testing results of the control performance in Table~\ref{tab:simple system}. Comparing the proposed Hybrid AoI with the MF Uniform baseline\footnote{We assume that the control reward of the MF Uniform method is always known despite the presence of packet dropout. This is an ideal case for the MF Uniform method.}, significant control performance improvements can be achieved. For example, the control performance has been improved by $70\%, 73\% \text{ and } 202\%$ in Hopper scenario 6, and InvDoublePen scenarios 2 and 6, respectively.
Even for the worst cases, i.e., HalfCheetah scenario 3 and 6, and InvDoublePen scenario 1, the improvement of $15\%, 28\% \text{ and } 27\%$ are obtained.
	
\textbf{Comparison between the AoI-based importance sampling and the uniform sampling methods.}
In Table~\ref{tab:simple system}, we compare the performance between the Hybrid AoI and the Hybrid Uniform algorithms over all scenarios. We see that improvements of $7\%-24\%$ can be obtained under scenarios with $5\%$ of uplink packet error rate, which further increases to $10\%-34\%$ under the $10\%$ uplink packet error rate, except for InvDoublePen scenario 1 with high packet dropout rates.
	
\subsubsection{Performance evaluation of the co-design algorithm in the high-mobility scenario} 
In Section~\ref{complex system}, we have proposed the joint estimator-controller-scheduler training algorithm with the consideration of the fading channel states and the communication costs. In particular, we propose to use the controller's Q-value for scheduler training.

\begin{figure}[t]
	\centering
	\includegraphics[width=0.49\textwidth, page=1]{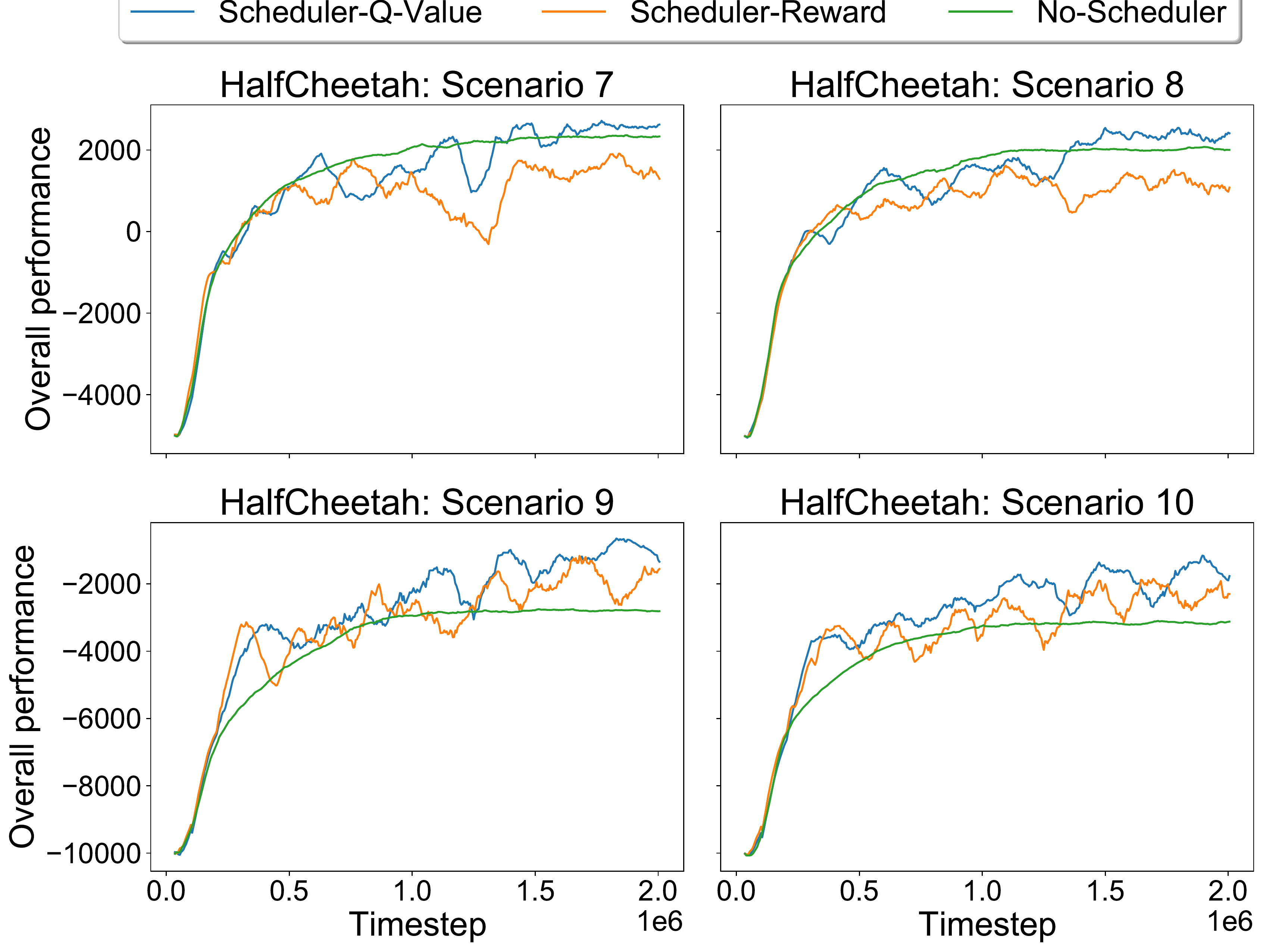}
	\caption{Comparison of learning curves in the high-mobility scenario.}
	\label{fig:learning curve3}
\end{figure}

For performance comparison, we consider two baselines in Table~\ref{tab:Compared methods 2}: ``No-Scheduler", where the sensor transmits in each time slot, and ``Scheduler-Reward", which use the conventional reward function for scheduler training.
We consider four WNCS settings with Markov fading channels in Table~\ref{tab:complex system}. 
The comparisons of the learning curves and the corresponding testing results are presented in Fig.~\ref{fig:learning curve3} and Table~\ref{tab:complex system}, respectively.

\begin{table}[t]
	\begin{threeparttable}
		\caption{Benchmark Policies (high-mobility scenario).}
		\label{tab:Compared methods 2}
		\setlength\tabcolsep{0pt} 
		
		\newcommand{\tabincell}[2]{\begin{tabular}{@{}#1@{}}#2\end{tabular}}  
		
		\begin{tabular*}{\columnwidth}{@{\extracolsep{\fill}} ll}
			\toprule
			\textbf{Method} & \textbf{Description} \\ 
			\midrule
			Scheduler-Q-Value (proposed) & \tabincell{l}{Scheduler trained by Q-value~\eqref{Q reward}} \\
			\midrule
			Scheduler-Reward & \tabincell{l}{Scheduler train by overall control reward~\eqref{total performance}} \\
			\midrule
			No-Scheduler & \tabincell{l}{No transmission scheduler} \\
			\bottomrule
		\end{tabular*}
		
		\smallskip
		\scriptsize
	\end{threeparttable}
\end{table}

\textbf{Performance evaluation over fading channels.} 
Considering slow and fast fading channel conditions $M_1,M_1$ and $M_2,M_2$, i.e., the left column and the right column in Fig.~\ref{fig:learning curve3}, the overall performance in the slow fading case is $9\%$ to $28\%$ higher than the fast fading one, with the low and high communication costs (rows 1 and 2 of in Fig.~\ref{fig:learning curve3}). This is because it is harder for both the controller and the transmission scheduler to learn adaptive policies with frequently changing channel states.
	
\textbf{Comparison with No-Scheduler.} 
In Table~\ref{tab:complex system}, comparing with the baseline No-Scheduler, the proposed Scheduler-Q-value algorithm increases the overall performance by about $11\%$ under communication cost $\bar{e}=5$, which further rises to $52\%$ with $\bar{e}=10$. It shows the effectiveness of the transmission scheduler in saving communication costs while guaranteeing control performance.
	
\textbf{Q-value vs. immediate reward.}
In Table~\ref{tab:complex system}, 
comparing the proposed Scheduler-Q-value method with the baseline Scheduler-Reward when training the scheduler, it can be observed that the proposed algorithm achieves a $72\%$-$99\%$ performance improvement over the baseline with the communication cost $\bar{e}=5$, and a $23\%$-$31\%$ improvement under communication cost $\bar{e}=10$. Therefore, using the long-term reward for scheduler training is much more effective than the short-term reward-based one.
Furthermore, it can be observed that the Scheduler-Q-value method leads to a smaller standard deviation than the baseline, indicating better training stability.

\section{Conclusions}\label{conclusion}
We have proposed a novel DL-based WNCS system with the awareness of both the AoI and the channel states.
We have then developed a novel DRL algorithm for joint controller and scheduler optimization utilizing both model-free and model-based data. In particular, data accuracy has been taken into account for enhancing learning efficiency.
We also develop novel schemes to stabilize the joint training of the controller and the scheduler. Our experiment results have demonstrated significant performance gain compared to the benchmarks.
For future works, we will consider multi-loop WNCSs over uplink and downlink wireless channels and investigate the co-design problems therein.

\bibliographystyle{IEEEtran}


\end{document}